\documentclass[journal]{IEEEtran}
\usepackage{amsmath,amsfonts}
\usepackage{array}

\usepackage{textcomp}
\usepackage{stfloats}
\usepackage{url}
\usepackage{verbatim}
\usepackage{subfigure}
\usepackage[linesnumbered,ruled,vlined]{algorithm2e}
\usepackage{cite}
\usepackage{amsmath,amssymb,amsfonts}
\usepackage{makecell}
\usepackage{graphicx}
\usepackage{textcomp}
\usepackage{bm}
\usepackage{diagbox}
\newtheorem{proposition}{\underline{Proposition}}
\newtheorem{remark}{\underline{Remark}}
\usepackage{xcolor}

\newcommand{\suo}{\hspace{-1pt}}
\def\BibTeX{{\rm B\kern-.05em{\sc i\kern-.025em b}\kern-.08em
		T\kern-.1667em\lower.7ex\hbox{E}\kern-.125emX}}
\usepackage{balance}
\usepackage{booktabs}
\allowdisplaybreaks[4]

\begin{document}
	
\title{Beyond Diagonal Intelligent Reflecting Surface Aided Integrated Sensing and Communication}
\author{Shuo Zheng,~\IEEEmembership{Graduate Student Member,~IEEE}, and Shuowen Zhang,~\IEEEmembership{Senior Member,~IEEE}
\thanks{This paper will be presented in part at the IEEE Workshop on Signal Processing and Artificial Intelligence for Wireless Communications (SPAWC), Surrey, United Kingdom, July 2025 \cite{SPAWCver}.
	
	The authors are with the Department of Electrical and Electronic Engineering, The Hong Kong Polytechnic University, Hong Kong SAR, China (e-mail: shuo.zheng@connect.polyu.hk; shuowen.zhang@polyu.edu.hk).}	}

%	\vspace{-20mm}
	
	\maketitle
	
%	\vspace{-10mm}
	
	\begin{abstract}
		
		  Beyond diagonal intelligent reflecting surface (BD-IRS) is a new promising IRS architecture for which the reflection matrix is not limited to the diagonal structure as for conventional IRS. In this paper, we study a BD-IRS aided uplink integrated sensing and communication (ISAC) system where sensing is performed in a device-based manner. Specifically, we aim to estimate the \textit{unknown} and \textit{random} location of an active target based on its uplink probing signals sent to a multi-antenna base station (BS) as well as the known prior distribution information  of the target's location. Multiple communication users also simultaneously send uplink signals, resulting in a challenging mutual interference issue  between sensing and communication. We first characterize the sensing performance metric by deriving the \textit{posterior Cram\'{e}r-Rao bound (PCRB)} of the mean-squared error (MSE) when prior information is available. Then, we formulate a BD-IRS reflection matrix optimization problem to maximize the minimum expected achievable rate among the multiple users subject to a constraint on the PCRB as well as the lossless and reciprocal constraints on the BD-IRS reflection matrix. The formulated problem is non-convex and challenging to solve. To tackle this problem, we propose a penalty dual decomposition (PDD) based algorithm which can find a high-quality suboptimal solution with polynomial-time complexity. In addition, we propose and optimize a time-division multiple access (TDMA) based scheme which removes the sensing-communication mutual interference. Numerical results verify the effectiveness of the proposed designs and provide useful design insights such as the optimal choice of multiple access scheme.
		 
	\end{abstract}
	\begin{IEEEkeywords}
		Beyond diagonal intelligent reflecting surface (BD-IRS), beyond diagonal reconfigurable intelligent surface (BD-RIS), integrated sensing and communication (ISAC), posterior Cram\'{e}r-Rao bound (PCRB).
	\end{IEEEkeywords}
	
	\section{Introduction}
	The sixth-generation (6G) wireless networks are expected to support a wide range of advanced applications, including autonomous driving, smart cities, and real-time monitoring. These applications often require both high-accuracy sensing and reliable communication. In this context, integrated sensing and communication (ISAC) has emerged as a promising technology, which goes beyond traditional communication and enables real-time collection, analysis, and processing of data from a variety of sensing anchors \cite{Liuzhang}. By integrating these functions, ISAC offers significant advantages in terms of spectrum and energy efficiency, as well as cost reduction. This approach not only improves sensing accuracy through the ubiquitous communication networks but also enhances network performance by utilizing the environmental information gathered from sensing, such as the locations of objects or obstacles. As a result, ISAC plays a crucial role in the evolution of 6G, and is considered as a key enabler of more intelligent and resource-efficient wireless networks \cite{ISACoverview,ISACoverview2}.
	
	 In recent years, intelligent reflecting surfaces (IRSs), also known as reconfigurable intelligent surfaces (RISs), and their diverse variants have gained increasing attention  \cite{tutorial,Roneone,Ronetwo,Rtwo}. An IRS is a reconfigurable metasurface composed of numerous reflecting elements capable of controlling signal propagation. By leveraging this capability, IRS can enhance communication reliability, reduce interference, and extend communication coverage, even in challenging environments. This technology opens the door to dynamic and controllable wireless propagation environments, leading to fully reconfigurable networks. By dynamically manipulating the wireless environment, IRS improves both communication and sensing capabilities \cite{IRScapacity,IRSOFDM,RISDoF,Liuzhang}, thereby augmenting ISAC systems \cite{IRSISACmag, IRSISACmag2}. Numerous research efforts have focused on IRS-aided ISAC systems \cite{IRSISAC1,IRSISAC3,IRSISAC6,BDIRSsensing,IRSISAC7}. For instance, \cite{IRSISAC1} investigated the joint waveform and discrete IRS phase shift design in ISAC systems. In \cite{IRSISAC3}, a heterogeneous 6G sensing system was proposed, which integrates active and passive anchors, with IRSs serving as cost-effective passive anchors that facilitate joint sensing with base stations (BSs) being the active anchors. In \cite{IRSISAC6}, physical-layer security in IRS-aided ISAC systems was studied, where active and passive beamforming were jointly designed to enhance ISAC performance while suppressing information leakage. \cite{BDIRSsensing} considered a device-based sensing scenario where an active target sends uplink signals to the BS with the help of an IRS. Furthermore, \cite{IRSISAC7} introduced a simultaneously transmitting and reflecting surface (STARS) enabled ISAC framework, where STARS divides space into distinct sensing and communication areas.
	
	However, the diagonal structure of the reflection matrix of conventional IRS limits its potential performance gains.  Recently, a novel IRS architecture, termed \emph{beyond diagonal IRS (BD-IRS)}\cite{BDIRSmag,BDIRS1}, also known as non-local IRS\cite{Rone,nonlocal} or beyond diagonal RIS (BD-RIS), has been proposed. Unlike diagonal IRS, BD-IRS allows inter-connections among its elements, thereby resulting in a beyond diagonal reflection matrix and offering greater design flexibility. 
	 By exploiting this property, BD-IRS can manipulate both the amplitude and phase of incident signals, which unlocks new opportunities for enhancing system performance. Thanks to the beyond diagonal structure of reflection matrix, BD-IRS is expected to outperform conventional IRS in many aspects.
		  To date, several studies have explored the design and optimization of BD-IRS in various scenarios \cite{BDIRS1,BDIRS3,BDIRS2,BDIRS4,BDIRS5,BDIRS7}. The concept and modeling of BD-IRS architecture were first introduced in \cite{BDIRS1} and then further studied in \cite{BDIRS3}.  \cite{BDIRS2} investigated the joint design of transmit precoders and BD-IRS reflection matrix to maximize the sum-rate in multi-user multiple-input-single-output (MISO) systems and showed BD-IRS's advantages over various architectures or operating modes. Despite its performance benefits, BD-IRS introduces additional circuit design complexity. To balance circuit complexity with performance, novel modeling approach, architecture design, and optimization framework for BD-IRS were proposed in \cite{BDIRS3} based on graph theory. Furthermore, the multi-sector BD-IRS concept was introduced in \cite{BDIRS4}, while \cite{BDIRS5} studied the modeling and optimization of BD-IRS with the existence of mutual coupling. In \cite{BDIRS7}, the BD-IRS reflection matrix was designed to minimize transmit power and maximize energy efficiency, subject to individual signal-to-interference-plus-noise ratio (SINR) constraints in a downlink multi-user MISO system. In \cite{Ronethree}, multiple BD-IRSs were exploited to assist the multiple access	system in a cooperative way. A distributed beamforming design with low information exchange overhead was proposed to improve the sum-rate. BD-IRS has also been explored for a wide range of applications, including rate splitting multiple access \cite{BDIRS9,BDIRSRSMA2}, unmanned aerial vehicles  \cite{BDIRSUAV}, mobile edge computing \cite{BDIRSMEC}, simultaneous wireless information and power transfer \cite{BDIRSSWIPT}, physical-layer security \cite{BDIRSSecur1}, and non-orthogonal multiple access  \cite{BDIRSNOMA}.
	
	In the context of ISAC, several studies have explored BD-IRS aided ISAC systems \cite{BDIRSISAC1,BDIRSISAC2,BDIRSISAC5,BDIRSISAC3,BDIRSISAC4,BDIRSISAC6}. In \cite{BDIRSISAC1}, the joint beamforming design in a downlink ISAC system was investigated to maximize throughput while satisfying a sensing signal-to-noise ratio (SNR) constraint. \cite{BDIRSISAC2} focused on the transmit power minimization problem under similar settings. In \cite{BDIRSISAC5}, a discrete phase setting was considered, and the maximization problem of the weighted sum of communication and sensing SNRs was studied. Additionally, \cite{BDIRSISAC3} leveraged BD-IRS operating in hybrid modes to enable full-space coverage. For multi-target scenarios, \cite{BDIRSISAC3} investigated a max-min signal-to-clutter-plus-noise ratio (SCNR) problem subject to communication rate constraints. \cite{BDIRSISAC1,BDIRSISAC2,BDIRSISAC5,BDIRSISAC3} utilized SNR or SCNR of the echo signal as the sensing performance metric, which, however, cannot reflect the error performance explicitly. Another widely recognized metric for sensing performance is the Cram\'{e}r-Rao bound (CRB), which provides an analytical lower bound for the mean-squared error (MSE). In \cite{BDIRSISAC4},  CRB was adopted to characterize the sensing performance of a multi-sector BD-IRS aided sensing system. Moreover, \cite{BDIRSISAC6} proposed a novel transmitter architecture for a BD-IRS aided millimeter wave ISAC system and  designed the joint beamforming to simultaneously maximize the sum-rate and minimize the largest eigenvalue of the CRB matrix. 
	
	However, CRB is only applicable to \textit{deterministic} parameters and is determined by their true values. In practice, the parameters to be estimated are often \textit{unknown} and \textit{random}, whose distributions can be known via statistical information \cite{xc, hky, dxm, xc2, yjyarxiv, wyz, xc3, yjy, hky2}. A new performance metric termed as \textit{posterior Cram\'{e}r-Rao bound (PCRB)} \cite{xc, hky, dxm, xc2, yjyarxiv, wyz, xc3, yjy, hky2} or Bayesian Cram\'{e}r-Rao bound (BCRB) \cite{BDIRSsensing,Kareem} serves as a lower bound on the MSE for sensing exploiting prior distribution information. Unlike CRB, PCRB is only a function of the prior distributions and does not depend on the true parameter values. Several prior studies have investigated PCRB minimization in ISAC systems \cite{xc, hky, dxm, xc2,yjyarxiv, wyz, xc3, yjy, hky2, Kareem, BDIRSsensing,zwf}. Nevertheless, how to optimize the performance of the BD-IRS aided ISAC or sensing systems with available prior distribution information of parameters still remains open for investigation, which motivates our study.
	
	In this paper, we investigate the BD-IRS reflection optimization in a BD-IRS aided multi-user uplink ISAC system, where a multiple-antenna BS simultaneously serves multiple single-antenna communication users in the uplink and senses the \textit{unknown} and \textit{random} location information of a target. The target is active and transmits probing signals to the BS for device-based sensing. The BS  senses the target's location based on the received signals and the known probability density function (PDF) of the target's location. There are three main challenges in this scenario. \emph{Firstly}, in downlink mono-static device-free sensing where the BS performs downlink communication and sensing based on the target-reflected echo signals, communication signals will not cause interference to sensing as they are known at the BS. However, in the studied uplink case, there exists \emph{mutual interference} between sensing and communication, as both the communication signals and target's location are \emph{unknown} at the BS. This thus calls for more powerful interference management via reflection optimization to strike an optimal balance between sensing and communication. \emph{Secondly}, the new general BD-IRS reflection matrix structure calls for new characterizations of the sensing performance and new optimization techniques as the results for conventional diagonal reflection matrices are not directly applicable. \emph{Thirdly}, how to judiciously design the BD-IRS reflection matrix to prioritize a range of possible locations with high probabilities under its structural constraints is a new challenging task.
	
	Motivated by the above challenges, we make the following contributions in this paper:
	\begin{itemize}
		\item Firstly, we characterize the PCRB for BD-IRS aided uplink ISAC as an explicit and tractable function of the BD-IRS reflection matrix, where the interference caused by communication is taken into account. We also derive a tractable lower bound of the expected achievable rate for each user averaged over the random interference caused by random target locations.
		\item Next, we formulate an optimization problem for the BD-IRS reflection matrix to maximize the minimum (approximate) expected communication rate among all users, while ensuring the sensing PCRB is below a threshold. This problem is highly non-convex and challenging due to the fractional expressions of communication rates and PCRB with mutual interference, as well as the structural constraints on the BD-IRS reflection matrix. To tackle this problem, we propose a penalty dual decomposition (PDD) \cite{penalty1} based algorithm to obtain a high-quality suboptimal solution with polynomial-time complexity. The proposed algorithm is also extended to the sensing-only case for PCRB minimization.
		\item Then, motivated by the potential limitation of the studied space-division multiple access (SDMA) scheme in mitigating severe interference, we propose a time-division multiple access (TDMA) scheme, where the optimal time allocation is derived in closed form.
		\item Finally, we provide extensive numerical results to validate the performance of the proposed scheme. The proposed algorithm is observed to converge fastly, and achieve superior performance compared with its conventional diagonal IRS counterparts and various benchmark schemes. Moreover, it is observed that TDMA may outperform SDMA in scenarios with heavy mutual interference between sensing and communication, e.g., when the user is located near the highly-probable locations of the target.
	\end{itemize}
	
	The remainder of this paper is organized as follows. Section \ref{sec2} presents the BD-IRS aided uplink ISAC system model. Section \ref{sec3} derives the PCRB to characterize the sensing performance. Section \ref{sec4} formulates the problem. Section \ref{sec5} proposes a suboptimal solution via a PDD-based algorithm. Section \ref{sec6.5} further proposes and optimizes a TDMA-based scheme. Numerical results are provided in Section \ref{sec6}. Finally, Section \ref{sec7} concludes this paper.
	
	\textit{Notations:} Vectors and matrices are denoted by boldface lower-case letters and boldface upper-case letters, respectively. $\mathbb{R}^{N\times M}$ and $\mathbb{C}^{N\times M}$ represent the space of $N\times M$ real matrices and $N\times M$ complex matrices, respectively. For a complex scalar $x$, $\vert x \vert$ and $\mathfrak{Re}\{x\}$ denote the absolute value and the real part, respectively. For a vector $\bm{x}$, $\Vert \bm{x}\Vert$ and $x_i$ denote the $l_2$ norm and the $i$-th entry, respectively.	For a matrix $\bm{X}$, $\bm{X}^T$, $\bm{X}^*$, $\bm{X}^H$, $\Vert \bm{X} \Vert_\mathrm{F}$, $\Vert \bm{X} \Vert_\infty$, ${X}_{i,j}$, $\mathrm{rank}(\bm{X})$, $\mathrm{vec}(\bm{X})$, and $\mathrm{vech}(\bm{X})$ denote the transpose, conjugate, conjugate transpose, Frobenius norm, infinity norm, $(i,j)$-th entry,  rank, vectorization, and half-vectorization, respectively. $\mathrm{vec}^{-1}(\bm{X})$ represents  inverse operation of vectorization such that $\mathrm{vec}^{-1}(\mathrm{vec}(\bm{X}))=\bm{X}$. $\mathrm{tr}(\bm{S})$, $\mathrm{det}(\bm{S})$, and $\bm{S}^{-1}$ represent the trace, determinant, and inverse of a non-singular matrix $\bm{S}$, respectively. The $M\times M$ identity matrix is denoted by $\bm{I}_M$.  $\mathrm{blkdiag}(\bm{X}_1,...,\bm{X}_N)$ refers to a block diagonal matrix with blocks $\bm{X}_1,...,\bm{X}_N$.  $j=\sqrt{-1}$ denotes the imaginary unit. $\mathcal{CN}(\bm{\mu},\bm{\Sigma})$ denotes the complex Gaussian distribution with mean vector $\bm{\mu}$ and covariance matrix $\bm{\Sigma}$; $\sim$ means ``distributed as". The first-order partial derivative is denoted by $\dot{\bm{f}}(x)=\frac{\partial \bm{f}(x)}{\partial x}$. $\mathcal{O}(\cdot)$ represents the standard big-O notation. $\mathbb{E}_{X}[\cdot]$ stands for the statistical expectation over a random variable $X$. The Kronecker product is denoted by $\otimes$. $a\  \mathrm{mod}\  b$ means $a$ modulo $b$.
	
	\section{System Model}\label{sec2}
	\begin{figure}[t]
		\centering
		\includegraphics[width=.9\linewidth]{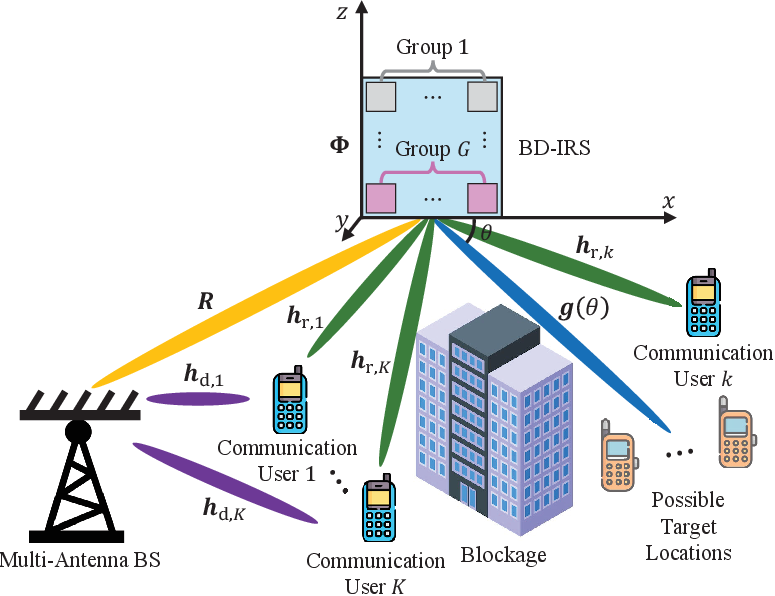}
		\vspace{-2mm}
		\caption{Illustration of an uplink BD-IRS aided ISAC system with prior distribution information.}
		\label{model}
%		\vspace{-.8cm}
		\vspace{-.5cm}
	\end{figure}
	We consider an uplink ISAC system, where a BS equipped with $N\geq 1$ receive antennas aims to sense the location information of an active target equipped with a single antenna and communicate with $K\geq 1$ single-antenna users via the uplink signals sent from the target/users and received at the BS. We focus on a challenging scenario where the direct link between the target and the BS is blocked by obstacles, and a BD-IRS equipped with $M\geq 1$ reflecting elements is deployed to create an additional reflected link for enabling target location sensing and enhancing uplink communication, as illustrated in Fig. \ref{model}. Specifically, we aim to sense the azimuth angle of the target with respect to the BD-IRS denoted by $\theta\in [0,\pi)$, as illustrated in Fig. \ref{model}, while the distance between the target and the BD-IRS is assumed to be known as $r$ meters (m).\footnote{The distance between the target and BD-IRS can be estimated via first estimating the distance from the target to the BS via the BD-IRS using time-of-arrival (ToA) methods (see, e.g., \cite{IRSISAC3,sqJSAC}) and then subtracting the known distance between the BS and the BD-IRS, or known \emph{a priori} based on target movement pattern or historic data \cite{xc,hky,dxm,xc2,yjyarxiv,wyz,xc3,yjy,hky2}.}  The value of $\theta$ is \emph{unknown}, while its PDF denoted by $p_{\Theta}(\theta)$ is known \emph{a priori} at the BS via historic data or target movement pattern \cite{xc,hky,dxm,xc2,yjyarxiv,wyz,xc3,yjy,hky2}.
		
	We consider a general reflective architecture at the BD-IRS, where the $M$ reflecting elements are grouped into $G$ groups, each consisting of $M_g$ elements, with $\sum_{g=1}^G M_g=M$. The elements in each group are inter-connected with a reflecting matrix $\bm{\Phi}_g\in \mathbb{C}^{M_g \times M_g}$, where $\bm{\Phi}_g^H\bm{\Phi}_g=\bm{I}_{M_g}$ and $\bm{\Phi}_g=\bm{\Phi}_g^T$ hold due to the lossless and reciprocal properties of BD-IRS \cite{BDIRS1}. The overall BD-IRS reflection matrix is given by
	\begin{align}
		\bm{\Phi}=\mathrm{blkdiag}\{\bm{\Phi}_1,...,\bm{\Phi}_G\}\label{Phi1}.
	\end{align}
	Note that when $G=1$, the above architecture corresponds to a \emph{fully-connected} BD-IRS where inter-connection exists between any two reflecting elements; when $G=M$, the above architecture reduces to a \emph{single-connected} BD-IRS, i.e., conventional IRS, where the BD-IRS reflection matrix becomes a diagonal matrix; while when $1<G<M$, the above architecture corresponds to a \emph{group-connected} BD-IRS, where inter-group elements are not connected. Moreover, we consider the case where the BD-IRS is located such that the channel from the target to the BD-IRS can be modeled as a line-of-sight (LoS) channel, which is denoted by $\bm{g}(\theta)\in \mathbb{C}^{M\times 1}$.\footnote{This is practically feasible with known PDF of the target's location.}
	Let $M_x$ and $M_z$ denote the number of reflecting elements in each row and column of the BD-IRS, respectively, with $M_xM_z=M$. Each element in $\bm{g}(\theta)$ is thus modeled as \cite{UPA}\footnote{In this paper, we put all the elements in each group on the same row \cite{rowgroup}, while our results are also directly applicable to other grouping strategies.}
	\begin{align}
		g_m(\theta) = \frac{\beta_0}{r}		 e^{j\frac{2\pi \Delta}{\lambda} (m-1)\mathrm{mod} M_x \cos \theta},\ m=1,...,M,
	\end{align}
	where $\beta_0$ denotes the reference channel gain at $1$ m, $\Delta$ denotes the spacing between adjacent reflecting elements in m, and $\lambda$ denotes the wavelength in m. Furthermore, let $\bm{R}\in \mathbb{C}^{N\times M}$ denote the channel from the BD-IRS to the BS. The effective target-BS channel is thus given by $\bm{R}\bm{\Phi} \bm{g}(\theta)$, for which the exact value is unknown due to the unknown $\theta$, while the statistical information can be derived based on the PDF of $\theta$. Let $\bm{h}_{\mathrm{r},k}\in \mathbb{C}^{M\times 1}$ denote the channel from user $k$ to the BD-IRS, and $\bm{h}_{\mathrm{d},k}\in \mathbb{C}^{N\times 1}$ denote the direct channel from user $k$ to the BS.\footnote{Note that this model is also applicable to the case where the user-BS link is blocked for some user(s), by setting the direct channel(s) as zero.} The effective channel from each communication user $k$ to the BS is thus given by 
	\begin{equation}
		\bm{h}_k(\bm{\Phi})=\bm{h}_{\mathrm{d},k}+\bm{R}\bm{\Phi}\bm{h}_{\mathrm{r},k},\  k=1,...,K.
	\end{equation}
	Note that the direct communication channels and cascaded communication channels can be estimated via various channel estimation techniques \cite{lhychannel, wrchannelarxiv, wrchannel}, and are assumed to be perfectly known in this paper.

We focus on ISAC in a block of $L$ symbol intervals during which both the target's location and the wireless channels remain static. Let $\sqrt{P_0}s_l\in \mathbb{C}$ denote the sensing signal sent from the target at each $l$-th symbol interval, where $P_0$ denotes the power budget at the target, and $s_l$ denotes the probing signal  \textit{known} at the BS. We further assume $|s_l|^2=1,\ l=1,...,L$, which yields a sample covariance for sensing of $\frac{1}{L}\sum_{l=1}^LP_0|s_l|^2=P_0$.\footnote{Note that in practice, there are various ways to design the sensing probing signals, e.g., via using the Zadoff-Chu sequence.}
 Let $c_{k,l}\sim \mathcal{CN}(0,1)$ denote the information symbol for each $k$-th communication user at the $l$-th symbol interval, and $P_k$ denote the average transmit power at each $k$-th communication user. The received signal at the BS in each $l$-th symbol interval is thus expressed as
	\begin{align}
		\!\!\!\bm{y}_l \!=\!\bm{R}\bm{\Phi}\bm{g}(\theta)\suo\sqrt{\suo P_0}s_l \!+\! \sum\limits_{k=1}^K \bm{h}_k(\bm{\Phi})\suo\sqrt{\suo P_k}c_{k,l} \!+\! \bm{n}_l,\  l\!=\!1,\suo...,\suo L,\!\!\!
		\label{BSsignal}
	\end{align}
	where $\bm{n}_l\sim \mathcal{CN}(\bm{0},\sigma^2\bm{I}_N)$ denotes the circularly symmetric complex Gaussian (CSCG) noise vector at the BS receiver with average noise power $\sigma^2$ at each antenna. The collection of the  received signals over $L$ symbol intervals denoted by $\bm{Y}=[\bm{y}_1,...,\bm{y}_L]\in\mathbb{C}^{N\times L}$ is thus given by
	\begin{align}
		\bm{Y}=\bm{R}\bm{\Phi}\bm{g}(\theta)\sqrt{P_0}\bm{s}^T + \sum\limits_{k=1}^K \bm{h}_k(\bm{\Phi})\sqrt{P_k}\bm{c}_{k}^T + \bm{N},
	\end{align}
	where $\bm{s}=[s_1,...,s_L]^T\in \mathbb{C}^{L\times 1}$ and $\bm{c}_k=[c_{k,1},...,c_{k,L}]^T\in\mathbb{C}^{L\times 1}$ denote the collections of sensing probing signals and information symbols from each user $k$, respectively;  $\bm{N}=[\bm{n}_1,...,\bm{n}_L]\in\mathbb{C}^{N\times L}$ denotes the collection of receiver noise. The BS performs sensing of $\theta$ based on $\bm{Y}$ and $p_\Theta(\theta)$, and detection of the information symbols sent at each $l$-th symbol interval based on each $\bm{y}_l$.
	
Specifically, at the BS receiver, linear receive  beamforming is adopted for information symbol detection. For each $k$-th user, denote $\bm{w}_k\in \mathbb{C}^{N\times 1}$ as the receive beamforming vector for user $k$, for which the beamforming output at each symbol interval is given by
\begin{align}
y_{k,l}
&=\bm{w}_k^H\bm{y}_l=\bm{w}_k^H\bm{h}_k(\bm{\Phi})\sqrt{P_k}c_{k,l}+\bm{w}_k^H\bm{R}\bm{\Phi}\bm{g}(\theta)\sqrt{P_0}s_l \nonumber\\
&\quad+\!\sum\limits_{\substack{k'=1\\k'\neq k}}^K  \bm{w}_k^H\bm{h}_{k'}(\bm{\Phi})\sqrt{P_{k'}}c_{k',l} \!+\! \bm{w}_k^H\bm{n}_l, \  l\!=\!1,...,L.\!\!\label{output}
\end{align}
Note that besides the interference from other communication users, the uplink signal sent from the sensing target also causes interference to each communication user, which cannot be canceled due to the unknown location of the target and consequently unknown channel $\bm{g}(\theta)$ from the target to the BD-IRS. The SINR for each $k$-th communication user is thus given by
\begin{align}
	&\!\!\!\gamma_k\suo(\suo\bm{w}_k\suo,\!\bm{\Phi}\suo,\suo\theta\suo)\!\suo
	=\!\suo\frac{P_{k} \vert \bm{w}_k^H \bm{h}_k(\bm{\Phi})\vert^2}{\sum\limits_{\substack{k'=1\\k' \neq k}}^K \!\!\suo P_{k'} \suo\vert \suo\bm{w}_k^H \suo\bm{h}_{k'}\suo(\suo\bm{\Phi}\suo) \suo\vert^2 \!\suo+\!\suo P_0 \vert \suo\bm{w}_k^H \suo\suo\bm{R}\bm{\Phi}\bm{g}\suo(\suo\theta\suo)\suo\vert^2 \!\suo+\!\suo \Vert \suo \bm{w}_k\suo\Vert^2\suo\sigma^2}\suo,\nonumber\\[-8pt]
	&\hspace{6.2cm} k=1,...,K.
\end{align}
Consequently, the achievable rate is given by $R_k(\bm{w}_k,\bm{\Phi},\theta)=\log_2(1+\gamma_k(\bm{w}_k,\bm{\Phi},\theta))$ in bits per second per Hertz (bps/Hz). Notice that the achievable rate for each communication user is determined by the angular location of the target, $\theta$, which is unknown. However, the known PDF of $\theta$ can be leveraged to calculate a tractable lower bound of the expected achievable rate over the random target locations, which is given by  
\begin{align}
	&\!\!\!\!\mathbb{E}_\theta[R_k(\bm{w}_k,\bm{\Phi},\theta)]\nonumber\\
	&\!\!\!\!\!\!\geq\!\suo
	\log_2\!\!\left(\!\! \suo1\!\!+\!\!\frac{P_{k} \vert \bm{w}_k^H \bm{h}_k(\bm{\Phi})\vert^2}{\!\sum\limits_{\substack{k'=1\\k' \neq k}}^K  \!\!\! P_{k'} \suo\vert\suo \bm{w}_k^H \suo\bm{h}_{k'}\suo(\suo\bm{\Phi}\suo)\suo \vert^2 \!\suo+\!\!  P_0 \suo\Vert\suo\bm{w}_k^H\suo \bm{R}\bm{\Phi}\bm{G}^{\frac{1}{2}}\suo\Vert^2  \!\suo+\!\suo \Vert\suo \bm{w}_k\suo\Vert^2\sigma^2}\!\!\suo\right)\!\!\!\!\!\label{jensen}\\
	&\!\!\!\!\!\!\triangleq \bar{R}_k(\bm{w}_k,\bm{\Phi}),\label{barR}
\end{align}
where $\bm{G}\triangleq\mathbb{E}_{\theta}[\bm{g}(\theta)\bm{g}^H(\theta)]=\int_{0}^\pi \bm{g}(\theta)\bm{g}(\theta)^Hd\theta$. Note that the inequality in \eqref{jensen} holds due to the convexity of $R_k(\bm{w}_k,\bm{\Phi},\theta)$ over $\bm{g}(\theta)\bm{g}(\theta)^H$ and Jensen's inequality. By noting that $\bar{R}_k(\bm{w}_k,\bm{\Phi})$ is only dependent on the design variables $\bm{w}_k$'s and $\bm{\Phi}$, we will adopt it as the communication performance metric in this paper, which serves as a lower bound of the expected achievable rate for each $k$-th user. Moreover, note that each receive beamforming vector $\bm{w}_k$ only affects the expected rate of its corresponding communication user, thus can be individually optimized to maximize $\bar{R}_k(\bm{w}_k,\bm{\Phi})$. To this end, we have the following proposition.

\begin{proposition}\label{pro1}
	The optimal receive beamforming vector for each $k$-th communication user that maximizes $\bar{R}_k(\bm{w}_k,\bm{\Phi})$ is given by
	\begin{align}
		\bm{w}_k^{\star} = \frac{\bm{\Sigma}_k^{-1}(\bm{\Phi})\bm{h}_k(\bm{\Phi})}{\Vert\bm{\Sigma}_k^{-1}(\bm{\Phi})\bm{h}_k(\bm{\Phi})\Vert}\label{wopt},
	\end{align}
	where $\bm{\Sigma}_k(\bm{\Phi}) = \sigma^2 \bm{I}_N+\sum_{k'=1,k' \neq k}^K P_{k'} \bm{h}_{k'}(\bm{\Phi})\bm{h}_{k'}^H(\bm{\Phi})
	+ P_0  \bm{R}\bm{\Phi}\bm{G} \bm{\Phi}^H\bm{R}^H
	$.
\end{proposition}
\begin{IEEEproof}
	Please refer to Appendix \ref{app1}.
\end{IEEEproof}

With the optimized $\bm{w}_k$ in Proposition \ref{pro1}, $\bar{R}_k(\bm{w}_k,\bm{\Phi})$ can be expressed as a function of only the BD-IRS reflection matrix $\bm{\Phi}$, which is given by
\begin{align}
	\bar{R}_k(\bm{\Phi}) = \log_2(1+ P_{k} \bm{h}_k^H(\bm{\Phi})\bm{\Sigma}_k^{-1}(\bm{\Phi})\bm{h}_k(\bm{\Phi})).
\end{align}

Note that the communication performance is critically dependent on the BD-IRS reflection matrix $\bm{\Phi}$. On the other hand, the design of $\bm{\Phi}$ also affects the sensing performance via the observations at the BS receiver, i.e., $\bm{Y}$. In the following, we will first characterize the sensing performance when both $\bm{Y}$ and $p_\Theta(\theta)$ are exploited. Then, we will formulate and study the BD-IRS reflection optimization problem to strike an optimal balance between sensing and communication.

\section{Sensing Performance Characterization via PCRB}\label{sec3}
Due to the difficulty in explicitly characterizing the MSE, we consider a tractable bound for the MSE as the sensing performance metric in this paper. Specifically, we will characterize the PCRB as a tractable lower bound of the sensing MSE when prior information about the distribution of $\theta$ can be exploited, which is tight in the moderate-to-high SNR regime.
To this end, we first derive the posterior Fisher information for estimating $\theta$, which is given by $F(\bm{\Phi})=F_\mathrm{O}(\bm{\Phi})+F_\mathrm{P}$ \cite{PCRBbook}. Specifically,  $F_{\mathrm{O}}(\bm{\Phi})$ represents the Fisher information from the observations in $\bm{Y}$, and $F_{\mathrm{P}}$ represents the Fisher information from the prior distribution information in $p_{\Theta}(\theta)$.
To derive $F_\mathrm{O}(\bm{\Phi})$, we first define $\bm{y}=\mathrm{vec}(\bm{Y})$ as follows:
\begin{align}
	\!\!\!\bm{y}\!=\!
	\begin{bmatrix}
		\suo\bm{R\Phi g}(\theta)\suo\sqrt{\suo P_0}s_1\suo\\
		\vdots\\
		\suo\bm{R\Phi g}(\theta)\suo\sqrt{\suo P_0}s_L\suo
	\end{bmatrix}
	\!\!+\!\!
	\begin{bmatrix}
		\sum_{k=1}^K \bm{h}_k(\bm{\Phi})\suo\sqrt{\suo P_k}c_{k,1}\\
		\vdots\\
		\sum_{k=1}^K \bm{h}_k(\bm{\Phi})\suo\sqrt{\suo P_k}c_{k,L}
	\end{bmatrix}\!\!+\!\!
	\begin{bmatrix}
		\suo\bm{n}_1\suo\\
		\vdots\\
		\suo\bm{n}_L\suo
	\end{bmatrix}\!\!.\!\!\!
\end{align}
Note that $\bm{y}\sim \mathcal{CN} (\bm{\mu}(\theta),\bm{C}(\theta))$, where $\bm{\mu}(\theta)=\sqrt{P_0}\bm{R}\bm{\Phi}\bm{g}(\theta) \otimes \bm{s}$ and 
	$\bm{C}(\theta) = \bm{\Sigma}_0(\bm{\Phi})\otimes \bm{I}_L$ with $\bm{\Sigma}_0(\bm{\Phi}) = \sum_{k=1}^K P_{k} \bm{h}_k(\bm{\Phi})\bm{h}_k^H(\bm{\Phi})+\sigma^2\bm{I}_N$ representing the covariance matrix of interference-plus-noise  for sensing. 
Then, the log-likelihood function for estimating $\theta$ from $\bm{y}$ is given by 
\begin{align}
	\!\!\!\ln(f(\bm{y}|\theta)) & \!=\! 2\sqrt{P_0}\mathfrak{Re}\Big\{\sum\limits_{l=1}^L\bm{y}_l^H \bm{\Sigma}_0^{-1}(\bm{\Phi}) \bm{R}\bm{\Phi}\bm{g}(\theta)s_l\Big\}\nonumber\\[-.3cm]
	&\quad \!-\!NL\ln(\pi \det(\bm{\Sigma}_0(\bm{\Phi})))
	 -\sum\limits_{l=1}^L\bm{y}_l^H\bm{\Sigma}_0^{-1}(\bm{\Phi})\bm{y}_l\nonumber\\[-.3cm]
	 &\quad \!-\!P_0\sum\limits_{l=1}^L (\bm{R}\bm{\Phi}\bm{g}(\theta)s_l)^H\bm{\Sigma}_0^{-1}(\bm{\Phi})\bm{R}\bm{\Phi}\bm{g}(\theta)s_l.\!\!
\end{align}
$F_\mathrm{O}(\bm{\Phi})$ can be thus derived as
\begin{align}
	 \!\!\! F_\mathrm{O}\suo(\suo\bm{\Phi}\suo)&\!=\!-\mathbb{E}_{\bm{y},\theta}\Big[ \frac{\partial^2 \ln (f(\bm{y}\vert \theta))}{\partial \theta^2}  \Big]\nonumber\\
	 &\!=\!2P_0\mathbb{E}_{\theta}\suo\Big[\suo\sum\limits_{l=1}^L (\bm{R\Phi}\dot{\bm{g}}(\theta)s_l)^H\bm{\Sigma}_0^{-1}(\bm{\Phi}) (\bm{R\Phi}\dot{\bm{g}}(\theta)s_l)\suo\Big]\!\\
	&\!=\!2P_0\suo\sum\limits_{l=1}^L \!\vert s_l\vert^2 \mathbb{E}_{\theta}\suo\Big[\suo (\bm{R\Phi}\dot{\bm{g}}(\theta))^H\bm{\Sigma}_0^{-1}\suo(\bm{\Phi}) (\bm{R\Phi}\dot{\bm{g}}(\theta))\suo\Big]\!\\
	&\!=\!2P_{0}L\mathbb{E}_{\theta} [(\bm{R\Phi}\dot{\bm{g}}(\theta))^H \bm{\Sigma}_0^{-1}(\bm{\Phi}) (\bm{R\Phi}\dot{\bm{g}}(\theta))],\label{Fo0}
\end{align}
where  $\dot{\bm{g}}(\theta)$ represents the derivative of $\bm{g}(\theta)$ with its $m$-th element being $\dot{g}_m(\theta) = -j\frac{2\pi \Delta}{\lambda}((m-1)\mathrm{mod} M_x\sin\theta) g_m(\theta),\ m=1,...,M$.

Note that the expression of $F_\mathrm{O}(\bm{\Phi})$ has a sophisticated form because of the complex involvement of the BD-IRS reflection matrix $\bm{\Phi}$. 
To further simplify the above expression, we define $\bm{U}\triangleq\mathbb{E}_{\theta}[\dot{\bm{g}}(\theta)\dot{\bm{g}}^H(\theta)]=\int_{0}^{\pi}\dot{\bm{g}}(\theta)\dot{\bm{g}}(\theta)^H d\theta$ and its eigenvalue decomposition (EVD) as $\bm{U}=\sum_{\zeta=1}^{R} \kappa_{\zeta}\bm{u}_{\zeta}\bm{u}_{\zeta}^H$ with $R=\mathrm{rank}(\bm{U})$ being its rank. Then, a more tractable expression of $F_\mathrm{O}(\bm{\Phi})$ can be derived as
\begin{align}
	\!\!\!\! F_\mathrm{O}\suo(\suo\bm{\Phi}\suo)&\!=\! 2P_0 L \bm{\varphi}^H\mathbb{E}_{\theta}\big[\suo\big(\suo (\suo\bm{R}^H\suo{\bm{\Sigma}}_0^{-1}\suo(\bm{\Phi})\suo\bm{R})^T
	\!\suo\otimes\! \dot{\bm{g}}(\theta)\dot{\bm{g}}^H\suo(\theta)\suo\big)\suo\big]\suo\bm{\varphi}\!\!\label{a}\\
	&\!=\!2P_0L \bm{\varphi}^H\suo\big(\suo(\suo\bm{R}^H\suo{\bm{\Sigma}}_0^{-1}\suo(\suo\bm{\Phi}\suo)\suo\bm{R})^T \!\suo\otimes\! (\mathbb{E}_{\theta}[\dot{\bm{g}}(\theta)\dot{\bm{g}}^H\suo(\theta)])\suo\big)\bm{\varphi}\!\suo\\
	&\!=\!2P_0L \bm{\varphi}^H\suo\big(\suo(\bm{R}^H{\bm{\Sigma}}_0^{-1}(\bm{\Phi})\bm{R})^T \!\suo\otimes\suo (\sum\limits_{\zeta=1}^{R} \kappa_{\zeta}\bm{u}_{\zeta}\bm{u}_{\zeta}^H)\big)\bm{\varphi}\\
&	\!=\!2P_0L \sum\limits_{\zeta=1}^{R}\suo \kappa_{\zeta}\bm{\varphi}^H\suo\big(\suo(\bm{R}^H\suo{\bm{\Sigma}}_0^{-1}\suo(\bm{\Phi})\bm{R})^T \!\suo\otimes\suo ( \bm{u}_{\zeta}\bm{u}_{\zeta}^H)\big)\bm{\varphi}\!\suo\\
	&\!=\! 2P_0L\sum\limits_{\zeta=1}^{R} \kappa_{\zeta}(\bm{R}\bm{\Phi}\bm{u}_{\zeta})^H \bm{\Sigma}_0^{-1}(\bm{\Phi}) (\bm{R}\bm{\Phi}\bm{u}_{\zeta}),\label{Fo1}
\end{align}
where $\bm{\varphi}=\mathrm{vec}(\bm{\Phi}^H)$, \eqref{a} holds due to $\bm{a}^H\bm{X}\bm{B}\bm{X}^H\bm{c}=\mathrm{vec}^H(\bm{X})(\bm{B}^T\otimes \bm{c}\bm{a}^H)\mathrm{vec}(\bm{X})$,
and \eqref{Fo1} holds due to $\mathrm{vec}^H(\bm{X})(\bm{B}^T\otimes \bm{c}\bm{a}^H)\mathrm{vec}(\bm{X})=\bm{a}^H\bm{X}\bm{B}\bm{X}^H\bm{c}$.

On the other hand, the value of $F_\mathrm{P}$ can be obtained as follows offline based on $p_\Theta(\theta)$: 
\begin{align}
\!\!\!\suo	F_{\mathrm{P}} \!=\!- \mathbb{E}_{\theta} \Big[\suo\frac{\partial^2\! \ln (p_{\Theta}(\theta))}{\partial \theta^2} \suo\Big]\!=\suo -\!\int_{-\infty}^{\infty} \! \frac{\partial^2 \!\ln (p_{\Theta}(\theta))}{\partial \theta^2}p_\Theta(\theta)d\theta.\!\!\suo
\end{align}

Based on $F_\mathrm{O}(\bm{\Phi})$ and $F_\mathrm{P}$, the PCRB for the MSE of sensing $\theta$ is given by
\begin{align}
	\!\!\!\!\!\mathrm{PCRB}_\theta(\suo\bm{\Phi}\suo)&\!=\!\frac{1}{F_{\mathrm{O}}(\bm{\Phi})\!+\!F_{\mathrm{P}}}\\
	&\!=\!\frac{1}{2P_0L\!\sum\limits_{\zeta=1}^{R}\!\suo \kappa_{\zeta}(\suo\bm{R}\bm{\Phi}\bm{u}_{\zeta}\suo)^H\suo \bm{\Sigma}_0^{-1}\suo(\suo\bm{\Phi}\suo) (\suo\bm{R}\bm{\Phi}\bm{u}_{\zeta}\suo)\!\suo+\!\suo F_\mathrm{P}}.\!\!\!\label{PCRB}
\end{align}

\begin{remark}[Practical Example of $p_\Theta(\theta)$]
 In practice, a typical example of the PDF $p_\Theta(\theta)$ is the Gaussian mixture model, where the realizations of $\theta$ tend to concentrate around several highly-probable angles, thus $p_\Theta(\theta)$ is the summation of multiple Gaussian PDFs. Let $\theta_i\in [0,\pi)$, $\sigma_i^2$, and $p_i$ denote the mean, variance, and weight of each $i$-th Gaussian PDF, $p_\Theta(\theta)$ can be expressed as $p_\Theta(\theta)=\sum_{i=1}^{I} p_{i}\frac{1}{\sqrt{2\pi}\sigma_{i}}e^{-\frac{(\theta-\theta_{i})^2}{2\sigma_{i}^2}}$. In this case, $F_\mathrm{P}$ can be derived as 
	$F_\mathrm{P}
	=\sum_{i=1}^{I}\frac{p_{i}}{\sigma_{i}^2}-\int_{-\infty}^\infty  \frac{\sum_{i_1=1}^{I}\sum_{i_2=1}^{I} \eta_{i_1}(\theta)\eta_{i_2}(\theta)(\frac{\theta-\theta_{i_1}}{\sigma_{i_1}^2}-\frac{\theta-\theta_{i_2}}{\sigma_{i_2}^2})^2}{2\sum_{i=1}^I \eta_{i}(\theta)}d\theta$,
where  $\eta_{i}(\theta)\triangleq p_{i}\frac{1}{\sqrt{2\pi}\sigma_{i}}e^{-\frac{(\theta-\theta_{i})^{2}}{2\sigma_{i}^{2}}}$ \cite{xc}. Note that when the variance of each Gaussian PDF goes to infinity, the Gaussian mixture model is reduced to the uniform distribution model; while when the variance of each Gaussian PDF tends to be zero, the Gaussian mixture model will approach a discrete model.
\end{remark}

\section{Problem Formulation}\label{sec4}
In this paper, we aim to optimize the BD-IRS reflection matrix to maximize the minimum (worst-case) expected achievable rate (approximated by its lower bound $\bar{R}_k(\bm{\Phi})$) among the multiple communication users, subject to a threshold on the sensing PCRB denoted by $\Gamma_{\mathrm{PCRB}}$. The optimization problem is formulated as
\begin{align}
	\!\!\mbox{(P1)}\  \max_{\bm{\Phi}}\min_{k=1,...,K}\,   &\log_2(1+ P_{k} \bm{h}_k^H(\bm{\Phi})\bm{\Sigma}_k^{-1}(\bm{\Phi})\bm{h}_k(\bm{\Phi}))
	\label{P1obj}\\
	\mathrm{s.t.}\ \,\   & \ \frac{1}{2 P_0  L\!\suo\sum\limits_{\zeta=1}^{R}\!\! \kappa_{\zeta}\suo(\suo\bm{R}\suo\bm{\Phi}\suo\bm{u}_{\zeta}\suo)^H\suo \bm{\Sigma}_0^{-1}\suo(\suo\bm{\Phi}\suo)\suo (\suo\bm{R}\suo\bm{\Phi}\suo\bm{u}_{\zeta}\suo)\!\suo+\!\suo F_\mathrm{P}}\nonumber\\
	&\hspace{3.8cm}\leq \Gamma_{\mathrm{PCRB}}\!\!\label{PCRBcons1}\\
	&\ \bm{\Phi}=\mathrm{blkdiag}\{\bm{\Phi}_1,...,\bm{\Phi}_G\}\label{blkdiag1}\\
	&\ \bm{\Phi}_g^H\bm{\Phi}_g = \bm{I}_{M_g},\  g=1,...,G\label{uni1}\\
	&\ \bm{\Phi}_g=\bm{\Phi}_g^T,\hspace{0.78cm} g=1,...,G.\label{sym1}
\end{align}

Note that Problem (P1) is a non-convex optimization problem as the objective function can be shown to be non-concave over $\bm{\Phi}$, the PCRB can be shown to be a non-convex function over $\bm{\Phi}$, and the constraints in \eqref{uni1} are non-convex. 

Moreover, the optimization of $\bm{\Phi}$ is challenging due to the generally conflicting goals of sensing and communication. For enhancing the communication performance, $\bm{\Phi}$ should be designed to achieve an optimal balance among the multi-user communication rates, while \emph{suppressing} the channel power from the highly-probable target angles to the BD-IRS to mitigate the interference caused by the target. On the other hand, for enhancing the sensing performance, $\bm{\Phi}$ should be designed towards \emph{increasing} the channel power from the highly-probable target angles to the BD-IRS, and mitigating the interference from all communication users via suppressing their channel powers. Note that in downlink mono-static ISAC where the BS performs sensing based on the target-reflected signals, the downlink communication signals contribute to sensing instead of causing interference. The \emph{mutual interference} between communication and sensing in the considered uplink ISAC system thus calls for a new judicious design of the BD-IRS reflection matrix to manage the mutual interference.

Finally, compared with uplink ISAC with conventional diagonal IRS (see, e.g., \cite{BDIRSsensing}), the BD-IRS reflection matrix is under more challenging constraints in \eqref{blkdiag1}-\eqref{sym1}, under which conventional techniques for simplifying the PCRB/rate expression and handling the optimization problem may not be directly applicable. However, it also provides new design degrees-of-freedom, the full exploitation of which may enable new performance gains.

In the next section, we will propose an efficient algorithm to tackle the non-convexity of (P1) by reformulating it into a more tractable equivalent form and applying PDD framework.

\section{Proposed Solution to Problem (P1)}\label{sec5}
First, by defining $\Gamma\triangleq\frac{\frac{1}{\Gamma_{\mathrm{PCRB}}}-F_\mathrm{P}}{2P_0L}$ and introducing an auxiliary variable $\alpha>0$ to represent the minimum SINR among the multiple users, Problem (P1) can be equivalently transformed into the following problem:
%\begin{align}
%	\mbox{(P2)}\  \min_{\alpha,\bm{\Phi}:(\suo\ref{blkdiag1}\suo)\suo,(\suo\ref{uni1}\suo)\suo,(\suo\ref{sym1}\suo)}\  & -\alpha
%	\label{P3obj}\\
%	\mathrm{s.t.}\quad\quad\ \  & \  P_k\bm{h}_k^H\suo(\suo\bm{\Phi}\suo)\suo\bm{\Sigma}_k^{-1}\suo(\suo\bm{\Phi}\suo)\bm{h}_k\suo(\suo\bm{\Phi}\suo)\!\geq \!\alpha,\  k\!=\!1,\suo...,\suo K\label{SINRcons3}\\
%	&\ \sum\limits_{\zeta=1}^{R} \kappa_{\zeta}(\bm{R}\bm{\Phi}\bm{u}_{\zeta})^H \bm{\Sigma}_0^{-1}(\bm{\Phi}) (\bm{R}\bm{\Phi}\bm{u}_{\zeta})\geq \Gamma.\label{PCRBcons3}
%\end{align}
\begin{align}
	\mbox{(P2)}\  \min_{\alpha,\bm{\Phi}}&\ -\alpha
	\label{P3obj}\\
	\mathrm{s.t.}&\  P_k\bm{h}_k^H(\bm{\Phi})\bm{\Sigma}_k^{-1}(\bm{\Phi})\bm{h}_k(\bm{\Phi})\geq \alpha,k=1,...,K\label{SINRcons3}\\
	&\ \sum\limits_{\zeta=1}^{R} \kappa_{\zeta}(\bm{R}\bm{\Phi}\bm{u}_{\zeta})^H \bm{\Sigma}_0^{-1}(\bm{\Phi}) (\bm{R}\bm{\Phi}\bm{u}_{\zeta})\geq \Gamma\label{PCRBcons3}\\
	&\ \bm{\Phi}=\mathrm{blkdiag}\{\bm{\Phi}_1,...,\bm{\Phi}_G\}\label{blkdiag3}\\
	&\ \bm{\Phi}_g^H\bm{\Phi}_g = \bm{I}_{M_g},\hspace{2.3cm}g=1,...,G\label{uni3}\\
	&\ \bm{\Phi}_g=\bm{\Phi}_g^T,\hspace{2.965cm}g=1,...,G.\label{sym3}
\end{align}

Then, we note that under the unitary constraints in \eqref{uni3}, optimization of $\bm{\Phi}$ for balancing the complicated PCRB and rate expressions is very challenging. To resolve this issue, we  introduce  a set of  auxiliary variables $\bm{\Psi}_g\in\mathbb{C}^{M_g\times M_g},\ g=1,...,G$, each satisfying $\bm{\Psi}_g^H\bm{\Psi}_g=\bm{I}_{M_g}$ and $\bm{\Phi}_g=\bm{\Psi}_g$. By this means, Problem (P2) and consequently Problem (P1) can be shown to be equivalent to the following problem:
%\begin{align}
%	\mbox{(P3)}\  \min_{\alpha,\bm{\Phi},\{\bm{\Psi}_g\}_{g=1}^G}\quad &-\alpha
%	\label{P4obj}\\
%	\mathrm{s.t.}\ \ \ \, \quad &\ P_k\bm{h}_k^H(\bm{\Phi})\bm{\Sigma}_k^{-1}(\bm{\Phi})\bm{h}_k(\bm{\Phi})\geq \alpha,\  k=1,...,K\label{SINRcons4}\\
%	&\  \sum\limits_{\zeta=1}^{R} \kappa_{\zeta}(\bm{R}\bm{\Phi}\bm{u}_{\zeta})^H \bm{\Sigma}_0^{-1}(\bm{\Phi}) (\bm{R}\bm{\Phi}\bm{u}_{\zeta})\geq \Gamma\label{PCRBcons4}\\
%	&\ \bm{\Phi}=\mathrm{blkdiag}\{\bm{\Phi}_1,...,\bm{\Phi}_G\}\label{blkdiag4}\\
%	&\ \bm{\Phi}_g=\bm{\Phi}_g^T,\ \hspace{3.4cm}  g=1,...,G\label{sym4}\\
%	&\ \bm{\Phi}_g=\bm{\Psi}_g,\ \hspace{3.45cm}  g=1,...,G\label{equal}\\
%	&\  \bm{\Psi}_g^H\bm{\Psi}_g=\bm{I}_{M_g},\ \hspace{2.63cm}  g=1,...,G\label{uni4}.
%\end{align}
\begin{align}
	\mbox{(P3)}\  \min_{\substack{\alpha,\bm{\Phi},\\ \{\bm{\Psi}_g\}_{g=1}^G}}&\ -\alpha
	\label{P4obj}\\
	\mathrm{s.t.}\ \,\,&\  P_k\bm{h}_k^H\!(\bm{\Phi})\bm{\Sigma}_k^{-1}\!(\bm{\Phi})\bm{h}_k(\bm{\Phi})\!\suo\geq\!\suo \alpha,k\!=\!1,...,K\label{SINRcons4}\\
	&\ \sum\limits_{\zeta=1}^{R} \kappa_{\zeta}(\bm{R}\bm{\Phi}\bm{u}_{\zeta})^H \bm{\Sigma}_0^{-1}(\bm{\Phi}) (\bm{R}\bm{\Phi}\bm{u}_{\zeta})\geq \Gamma\label{PCRBcons4}\!\!\\
	&\ \bm{\Phi}=\mathrm{blkdiag}\{\bm{\Phi}_1,...,\bm{\Phi}_G\}\label{blkdiag4}\\
	&\ \bm{\Phi}_g=\bm{\Phi}_g^T,\hspace{2.64cm}g\!=\!1,...,G\label{sym4}\\
	&\ \bm{\Phi}_g=\bm{\Psi}_g,\hspace{2.67cm}g\!=\!1,...,G\label{equal}\\
	&\  \bm{\Psi}_g^H\bm{\Psi}_g=\bm{I}_{M_g},\hspace{1.94cm}g\!=\!1,...,G\label{uni4}.\!
\end{align}
Note that without the constraints in \eqref{equal}, the optimization of $\bm{\Phi}$ will be more tractable. Motivated by this, we propose a PDD-based algorithm to decouple $\bm{\Phi}_g$ and $\bm{\Psi}_g$ in \eqref{equal}. Specifically, by transforming the constraints in \eqref{equal} into a penalty term, the augmented Lagrange (AL) function can be written as
\begin{align}
&\!\!\!\!\mathcal{L}(\alpha,\{\bm{\Phi}_g\}_{g=1}^G,\{\bm{\Psi}_g\}_{g=1}^G,\{\bm{\Lambda}_g\}_{g=1}^G)\nonumber\\
&\!\!\!\!\!\!=\!-\alpha\!+\!\!\sum\limits_{g=1}^G\!\mathfrak{Re}\big\{\!\mathrm{Tr}\big(\bm{\Lambda}_g^H(\bm{\Phi}_g\!-\!\bm{\Psi}_g)\big)\!\big\}\!\!+\!\!\sum\limits_{g=1}^G\!\frac{1}{2\rho}\Vert \bm{\Phi}_g\!-\!\bm{\Psi}_g \Vert^2_\mathrm{F},\!\!\!\!
\end{align}
where $\bm{\Lambda}_g\in\mathbb{C}^{M_g\times M_g},\ g=1,...,G$ denote the set of dual variables associated with the constraints in \eqref{equal}, and $\rho$ denotes the penalty parameter.
Then, we formulate the AL problem as
%\begin{align}
%	\mbox{(P4)}\  \min_{\alpha,\bm{\Phi}, \{\bm{\Psi}_g\}_{g=1}^G}& -\alpha+\sum\limits_{g=1}^G\mathfrak{Re}\big\{\mathrm{Tr}\big(\bm{\Lambda}_g^H(\bm{\Phi}_g-\bm{\Psi}_g)\big)\big\}+\sum\limits_{g=1}^G\frac{1}{2\rho}\Vert \bm{\Phi}_g-\bm{\Psi}_g \Vert^2_\mathrm{F}
%	\label{P5obj}\\
%	\mathrm{s.t.}\ \,\ \  &\  P_k\bm{h}_k^H(\bm{\Phi})\bm{\Sigma}_k^{-1}(\bm{\Phi})\bm{h}_k(\bm{\Phi})\geq \alpha,\  k=1,...,K\label{SINRcons5}\\
%	&\ \sum\limits_{\zeta=1}^{R} \kappa_{\zeta}(\bm{R}\bm{\Phi}\bm{u}_{\zeta})^H \bm{\Sigma}_0^{-1}(\bm{\Phi}) (\bm{R}\bm{\Phi}\bm{u}_{\zeta})\geq \Gamma\label{PCRBcons5}\\
%	&\ \bm{\Phi}=\mathrm{blkdiag}\{\bm{\Phi}_1,...,\bm{\Phi}_G\}\label{blkdiag5}\\
%	&\  \bm{\Phi}_g=\bm{\Phi}_g^T,\hspace{3.4cm}  g=1,...,G\label{sym5}\\
%	&\  \bm{\Psi}_g^H\bm{\Psi}_g=\bm{I}_{M_g}, \hspace{2.63cm} g=1,...,G\label{uni5}.
%\end{align}
\begin{align}
	\mbox{(P4)}\  \min_{\substack{\alpha,\bm{\Phi},\\ \{\bm{\Psi}_g\}_{g=1}^G}}&\ -\alpha+\sum\limits_{g=1}^G\mathfrak{Re}\big\{\mathrm{Tr}\big(\bm{\Lambda}_g^H(\bm{\Phi}_g-\bm{\Psi}_g)\big)\big\}\nonumber\\
	&\hspace{1cm}+\sum\limits_{g=1}^G\frac{1}{2\rho}\Vert \bm{\Phi}_g-\bm{\Psi}_g \Vert^2_\mathrm{F}
	\label{P5obj}\\
	\mathrm{s.t.}\ \,\,&\  P_k\bm{h}_k^H\!(\bm{\Phi})\bm{\Sigma}_k^{-1}\!(\bm{\Phi})\bm{h}_k(\bm{\Phi})\!\suo\geq\!\suo \alpha,\  k\!=\!1,...,K\!\!\label{SINRcons5}\\
	&\ \sum\limits_{\zeta=1}^{R} \kappa_{\zeta}(\bm{R}\bm{\Phi}\bm{u}_{\zeta})^H \bm{\Sigma}_0^{-1}(\bm{\Phi}) (\bm{R}\bm{\Phi}\bm{u}_{\zeta})\geq \Gamma\label{PCRBcons5}\\
	&\ \bm{\Phi}=\mathrm{blkdiag}\{\bm{\Phi}_1,...,\bm{\Phi}_G\}\label{blkdiag5}\\
	&\  \bm{\Phi}_g=\bm{\Phi}_g^T,\  \hspace{2.66cm}g\!=\!1,...,G\!\!\!\!\label{sym5}\\
	&\  \bm{\Psi}_g^H\bm{\Psi}_g=\bm{I}_{M_g},\   \hspace{1.96cm}g\!=\!1,...,G \!\!\label{uni5}.\!\!\!
\end{align}

To solve the AL problem, we iteratively optimize $\bm{\Phi}$ and $\{\bm{\Psi}_g\}_{g=1}^G$ in the inner loop of PDD, while the outer loop selectively updates the dual variables or the penalty parameter. The optimization procedure is given in detail below.

\vspace{-.3cm}
\subsection{(Inner Loop) Optimization of $\bm{\Phi}$ with Given Auxiliary Variables $\{\bm{\Psi}_g\}_{g=1}^G$}
With given $\{\bm{\Psi}_g\}_{g=1}^G$, (P4) reduces to the following sub-problem:
%\begin{align}
%	\mbox{(P4-I)}\quad \min_{\alpha,\bm{\Phi}}\quad & -\alpha+\sum\limits_{g=1}^G\mathfrak{Re}\big\{\mathrm{Tr}\big(\bm{\Lambda}_g^H(\bm{\Phi}_g-\bm{\Psi}_g)\big)\big\}+\sum\limits_{g=1}^G\frac{1}{2\rho}\Vert \bm{\Phi}_g-\bm{\Psi}_g \Vert^2_\mathrm{F}
%	\label{P6obj}\\
%	\mathrm{s.t.}\quad &\  P_k\bm{h}_k^H(\bm{\Phi})\bm{\Sigma}_k^{-1}(\bm{\Phi})\bm{h}_k(\bm{\Phi})\geq \alpha,\  k=1,...,K\label{SINRcons6}\\
%	&\  \sum\limits_{\zeta=1}^{R} \kappa_{\zeta}(\bm{R}\bm{\Phi}\bm{u}_{\zeta})^H \bm{\Sigma}_0^{-1}(\bm{\Phi}) (\bm{R}\bm{\Phi}\bm{u}_{\zeta})\geq \Gamma\label{PCRBcons6}\\
%	&\  \bm{\Phi}=\mathrm{blkdiag}\{\bm{\Phi}_1,...,\bm{\Phi}_G\}\label{blkdiag6}\\
%	&\   \bm{\Phi}_g=\bm{\Phi}_g^T,\hspace{3.47cm}\, g=1,...,G\label{sym6}.
%\end{align}
\begin{align}
	\!\!\!\!\mbox{(P4-I)}\,  \min_{\alpha,\bm{\Phi}}\ & -\!\alpha\!+\!\!\sum\limits_{g=1}^G\!\mathfrak{Re}\suo\big\{\!\mathrm{Tr}\suo\big(\!\bm{\Lambda}_g^H\!(\!\bm{\Phi}_g\!\!-\!\!\bm{\Psi}_g)\!\big)\!\big\}\!\!+\!\!\sum\limits_{g=1}^G\!\frac{1}{2\rho}\!\Vert \bm{\Phi}_g\!\!-\!\!\bm{\Psi}_g \Vert^2_\mathrm{F}
	\label{P6obj}\\
	\mathrm{s.t.}\ &  P_k\bm{h}_k^H(\bm{\Phi})\bm{\Sigma}_k^{-1}(\bm{\Phi})\bm{h}_k(\bm{\Phi})\geq \alpha,k=1,...,K\label{SINRcons6}\\
	& \sum\limits_{\zeta=1}^{R} \kappa_{\zeta}(\bm{R}\bm{\Phi}\bm{u}_{\zeta})^H \bm{\Sigma}_0^{-1}(\bm{\Phi}) (\bm{R}\bm{\Phi}\bm{u}_{\zeta})\geq \Gamma\label{PCRBcons6}\\
	& \bm{\Phi}=\mathrm{blkdiag}\{\bm{\Phi}_1,...,\bm{\Phi}_G\}\label{blkdiag6}\\
	&  \bm{\Phi}_g=\bm{\Phi}_g^T,\hspace{2.95cm}g=1,...,G\label{sym6}.
\end{align}

Note that (P4-I) is still a non-convex problem due to the non-convex constraints in \eqref{SINRcons6} and \eqref{PCRBcons6}. Moreover, the involvement of the inversions of $\bm{\Sigma}_k(\bm{\Phi})$'s and $\bm{\Sigma}_0(\bm{\Phi})$ makes (P4-I) more challenging. To tackle this issue, we transform (P4-I) into an equivalent form in the following proposition.

\begin{proposition}\label{pro2}
	Problem (P4-I) is equivalent to the following problem with auxiliary variables $\bm{\nu}_{\zeta}\in\mathbb{C}^{N\times 1},\zeta=1,...,R$ and $\bm{\nu}_k\in\mathbb{C}^{N\times 1},k=1,...,K$:
%\begin{align}
%	\mbox{(P4-I-eqv)}\min_{\substack{\alpha,\bm{\Phi},\{\bm{\nu}_k\}_{k=1}^K\\ \{\bm{\nu}_{\zeta}\}_{\zeta=1}^{R}}}\quad & -\alpha+\sum\limits_{g=1}^G\mathfrak{Re}\big\{\mathrm{Tr}\big(\bm{\Lambda}_g^H(\bm{\Phi}_g-\bm{\Psi}_g)\big)\big\}+\sum\limits_{g=1}^G\frac{1}{2\rho}\Vert \bm{\Phi}_g-\bm{\Psi}_g \Vert^2_\mathrm{F}
%	\label{P7obj}\\
%	\mathrm{s.t.}\quad\quad\, &\  f_k(\bm{\Phi},\bm{\nu}_k)+\frac{\alpha}{P_{k}}\leq 0,\   k=1,...,K\label{SINRcons7}\\
%	& \ \sum\limits_{\zeta=1}^{R} \kappa_{\zeta} f_{\zeta}(\bm{\Phi},\bm{\nu}_{\zeta})+\Gamma\leq 0\label{PCRBcons7}\\
%	&\  \bm{\Phi}=\mathrm{blkdiag}\{\bm{\Phi}_1,...,\bm{\Phi}_G\}\label{blkdiag7}\\
%	& \  \bm{\Phi}_g=\bm{\Phi}_g^T,\hspace{1.92cm} g=1,...,G\label{sym7},
%\end{align}
\begin{align}
	\!\!\mbox{(P4-I-eqv)}\min_{\substack{\alpha,\bm{\Phi},\{\bm{\nu}_k\}_{k=1}^K,\\ \{\bm{\nu}_{\zeta}\}_{\zeta=1}^{R}}}& \, -\alpha+\sum\limits_{g=1}^G\mathfrak{Re}\big\{\mathrm{Tr}\big(\bm{\Lambda}_g^H(\bm{\Phi}_g-\bm{\Psi}_g)\big)\big\}\nonumber\\[-.2cm]
	&\hspace{1cm}+\sum\limits_{g=1}^G\frac{1}{2\rho}\Vert \bm{\Phi}_g-\bm{\Psi}_g \Vert^2_\mathrm{F}
	\label{P7obj}\\
	\mathrm{s.t.}\quad\  & f_k(\bm{\Phi},\bm{\nu}_k)\!+\!\frac{\alpha}{P_{k}}\leq 0, k=1,...,K\label{SINRcons7}\!\!\\[-.1cm]
	& \sum\limits_{\zeta=1}^{R} \kappa_{\zeta} f_{\zeta}(\bm{\Phi},\bm{\nu}_{\zeta})+\Gamma\leq 0\label{PCRBcons7}\\
	& \bm{\Phi}=\mathrm{blkdiag}\{\bm{\Phi}_1,...,\bm{\Phi}_G\}\label{blkdiag7}\\
	&  \bm{\Phi}_g=\bm{\Phi}_g^T,\hspace{1.45cm}g=1,...,G\label{sym7},\!\!\!
\end{align}
where $	f_{\zeta}(\bm{\Phi},\bm{\nu}_{\zeta})\triangleq
\bm{\nu}_{\zeta}^H\bm{\Sigma}_0(\bm{\Phi})\bm{\nu}_{\zeta}-2\mathfrak{Re}\{\bm{\nu}_{\zeta}^H\bm{R\Phi u}_{\zeta}\},\ \zeta=1,...,R$ and
	$f_k(\bm{\Phi},\bm{\nu}_k)\triangleq
	\bm{\nu}_k^H\bm{\Sigma}_k(\bm{\Phi})\bm{\nu}_k-2\mathfrak{Re}\{\bm{\nu}_k^H\bm{ h}_k\},\ k=1,...,K$.
\end{proposition}

\begin{IEEEproof}
	Please refer to Appendix \ref{app2}.
\end{IEEEproof}

Although Problem (P4-I-eqv) is still non-convex, we propose an alternating optimization algorithm to iteratively optimize $(\alpha,\bm{\Phi})$ and $(\{\bm{\nu}_{\zeta}\}_{\zeta=1}^{R}, \{\bm{\nu}_k\}_{k=1}^K)$ as follows. 

\subsubsection{Optimization of $(\{\bm{\nu}_{\zeta}\}_{\zeta=1}^{R}, \{\bm{\nu}_k\}_{k=1}^K)$ with Given $(\alpha,\bm{\Phi})$} 
With given $(\alpha,\bm{\Phi})$, the optimal $\{\bm{\nu}_{\zeta}\}_{\zeta=1}^{R}$ and $\{\bm{\nu}_k\}_{k=1}^K$ can be derived below as shown in the proof of Proposition \ref{pro2}:
\begin{align}
	\bm{\nu}_{\zeta}^{\star} &= \bm{\Sigma}_0^{-1}(\bm{\Phi})\bm{R\Phi u}_{\zeta},\  \zeta=1,...,R\label{lambdar}\\
	\bm{\nu}_k^{\star} &= \bm{\Sigma}_k^{-1}(\bm{\Phi})\bm{h}_k(\bm{\Phi}),\  \hspace{0.018cm} k=1,...,K.\label{lambdak}
\end{align}
\subsubsection{Optimization of $(\alpha,\bm{\Phi})$ with Given $(\{\bm{\nu}_{\zeta}\}_{\zeta=1}^{R}, \{\bm{\nu}_k\}_{k=1}^K)$}

With given $(\{\bm{\nu}_{\zeta}\}_{\zeta=1}^{R}, \{\bm{\nu}_k\}_{k=1}^K)$, a key difficulty in efficiently optimizing $(\alpha,\bm{\Phi})$ lies in the complex expression of  $f_{\zeta}(\bm{\Phi},\bm{\nu}_{\zeta})$ and $f_{k}(\bm{\Phi},\bm{\nu}_{k})$'s with respect to matrix $\bm{\Phi}$. Moreover, the large number of optimization variables in $\bm{\Phi}$ also hinders the development of a low-complexity algorithm. This  motivates us to present an equivalent lower-dimension replacement of $\bm{\Phi}$ yielding a more tractable form of $f_{\zeta}(\bm{\Phi},\bm{\nu}_{\zeta})$ and $f_{k}(\bm{\Phi},\bm{\nu}_{k})$'s. Specifically, define $\bm{\varphi}_g\triangleq\mathrm{vec}(\bm{\Phi}_g),\ g=1,...,G$. Then, $\bm{\Phi}$ can be expressed as $\bm{\Phi}=\mathrm{vec}^{-1}(\sum_{g=1}^G \bm{Q}_g \bm{\varphi}_g)$, where $\bm{Q}_g\in\{0,1\}^{M^2\times M_g^2},\ g=1,...,G$ denote a set of indexing matrices given in Appendix \ref{Qg}. Furthermore, we reduce the dimension of $\bm{\varphi}_g$'s by exploiting the reciprocal property of $\bm{\Phi}$. Define $\bm{\phi}_g\triangleq\mathrm{vech}(\bm{\Phi}_g)\in \mathbb{C}^{\frac{M_g(M_g+1)}{2}\times 1}$. By introducing a set of duplication matrices $\bm{D}_g\in \{0,1\}^{M_g^2\times \frac{M_g(M_g+1)}{2}},\ g=1,...,G$ whose construction is provided in Appendix \ref{Dg}, we have $\bm{D}_g\bm{\phi}_g=\bm{\varphi}_g,\ g=1,...,G$ and consequently $
\bm{\Phi}_g = \mathrm{vec}^{-1}(\bm{D}_g\bm{\phi}_g),\ g=1,...,G$. The BD-IRS reflection matrix $\bm{\Phi}$ with $M^2$ elements can be equivalently represented by $\{\bm{\phi}_g\}_{g=1}^G$ with $\sum_{g=1}^G \frac{M_g(M_g+1)}{2}$ elements as follows:
\begin{equation}
	\bm{\Phi}=\mathrm{vec}^{-1}\left(\sum_{g=1}^G \bm{Q}_g \bm{D}_g\bm{\phi}_g\right).\label{recover}
\end{equation}
As a result, the optimization of $\{\bm{\phi}_g\}_{g=1}^G$ is equivalent to the optimization of $\bm{\Phi}$. 

Then, we simplify  (P4-I-eqv) based on $\{\bm{\phi}_g\}_{g=1}^G$. To this end, we first expand $f_{\zeta}(\bm{\Phi},\bm{\nu}_{\zeta})$ as
\begin{align}
	&\!\!\!\!f_{\zeta}(\bm{\Phi},\bm{\nu}_{\zeta})=\bm{\nu}_{\zeta}^H\bm{\Sigma}_0(\bm{\Phi}) \bm{\nu}_{\zeta}-2\mathfrak{Re}\{\bm{\nu}_{\zeta}^H\bm{R\Phi u}_{\zeta}\}\\
	&\!\!\!\!\!\!=\!\sum\limits_{k=1}^K\! P_{k} \bm{\nu}_{\zeta}^H\! \bm{R\Phi h}_{\mathrm{r},k}   \bm{h}_{\mathrm{r},k} ^H \!\bm{\Phi}^H\!\!\bm{R}^H\!\bm{\nu}_{\zeta}\!\suo+\suo\!\sum\limits_{k=1}^K \!P_{k} \vert\bm{\nu}_{\zeta}^H\! \bm{h}_{\mathrm{d},k}   \vert^2\!\!+\!\sigma^2\Vert\bm{\nu}_{\zeta}\Vert^2\nonumber\\
	&\!\!\!\!+\!2\sum\limits_{k=1}^K \!P_{k} \mathfrak{Re}\{\bm{\nu}_{\zeta}^H \bm{R\Phi h}_{\mathrm{r},k}  \bm{h}_{\mathrm{d},k} ^H \bm{\nu}_{\zeta}\}\!-\!2\mathfrak{Re}\{\bm{\nu}_{\zeta}^H\bm{R\Phi u}_{\zeta}\}\label{fropt1}.
\end{align}

Based on (\ref{recover}), the first term of (\ref{fropt1}) can be re-expressed as
\begin{align}
	&\sum\limits_{k=1}^K P_{k} \bm{\nu}_{\zeta}^H \bm{R\Phi h}_{\mathrm{r},k}  \bm{h}_{\mathrm{r},k}^H \bm{\Phi}^H\bm{R}^H\bm{\nu}_{\zeta}\nonumber\\
	&=\sum\limits_{k=1}^K P_{k} \mathrm{vec}(\bm{\nu}_{\zeta}^H \bm{R\Phi h}_{\mathrm{r},k}  \bm{h}_{\mathrm{r},k}^H \bm{\Phi}^H\bm{R}^H\bm{\nu}_{\zeta})\label{aa}\\
	&=\sum\limits_{k=1}^K P_{k} \mathrm{vec}^H(\bm{\Phi}) \underbrace{(\bm{h}_{\mathrm{r},k}^*\bm{h}_{\mathrm{r},k}^T\otimes \bm{R}^H \bm{\nu}_{\zeta} \bm{\nu}_{\zeta}^H \bm{R})}_{\bm{V}_{\zeta,k}}\mathrm{vec}(\bm{\Phi}) \label{bb}\\[-.15cm]
	&= \sum\limits_{k=1}^K P_{k} \Big(\sum_{g=1}^G \bm{Q}_g \bm{D}_g\bm{\phi}_g\Big)^H \bm{V}_{\zeta,k}\Big(\sum_{g=1}^G \bm{Q}_g \bm{D}_g\bm{\phi}_g\Big) \\
	&= \Big(\sum_{g=1}^G \bm{Q}_g \bm{D}_g\bm{\phi}_g\Big)^H\bm{V}_{\zeta}\Big(\sum_{g=1}^G \bm{Q}_g \bm{D}_g\bm{\phi}_g\Big),\label{fropt1vec}
\end{align}
where \eqref{aa} holds since $\bm{\nu}_{\zeta}^H \bm{R\Phi h}_k  \bm{h}_k^H(\bm{\Phi}) \bm{\Phi}^H\bm{R}^H\bm{\nu}_{\zeta}$ is a scalar; \eqref{bb} holds due to $\bm{a}^H\bm{X}\bm{B}\bm{X}^H\bm{c}=\mathrm{vec}^H(\bm{X})(\bm{B}^T\otimes \bm{c}\bm{a}^H)\mathrm{vec}(\bm{X})$ and $\bm{V}_{\zeta}=\sum_{k=1}^K P_{k} \bm{V}_{\zeta,k}$. Note that $\bm{V}_{\zeta,k}$ is a positive semi-definite (PSD) matrix because it is the Kronecker product of PSD matrices \cite{kronecker}. Thus, $\bm{V}_{\zeta}$ is also a PSD matrix. The fourth term of \eqref{fropt1} can be re-expressed as
\begin{align}
	&\!\!\!2\sum\limits_{k=1}^K P_{k}	\mathfrak{Re}\{ \bm{\nu}_{\zeta}^H \bm{R\Phi h}_{\mathrm{r},k}  \bm{h}_{\mathrm{d},k} ^H  \bm{\nu}_{\zeta}\}\nonumber\\
	&\!\!\!\!=\!2\sum\limits_{k=1}^K P_{k}\mathfrak{Re}\{\mathrm{Tr}(\bm{h}_{\mathrm{r},k}  \bm{h}_{\mathrm{d},k}^H  \bm{\nu}_{\zeta}\bm{\nu}_{\zeta}^H\bm{R\Phi})\}\\
	 &\!\!\!\!=\!2\sum\limits_{k=1}^K \! P_{k}\mathfrak{Re}\suo\Big\{\!\mathrm{vec}^H\suo(\suo\bm{R}^H \suo\bm{\nu}_{\zeta} \bm{\nu}_{\zeta}^H\suo\bm{h}_{\mathrm{d},k}\bm{h}_{\mathrm{r},k}^H)\suo\Big(\!\sum_{g=1}^G \!\bm{Q}_g \bm{D}_g\bm{\phi}_g\suo\Big)\suo\Big\},\!\!\!\label{fropt3vec}
	 \end{align}
	 which holds due to $\mathrm{Tr} (\bm{A}^H \bm{B})=\mathrm{vec}^H(\bm{A})\mathrm{vec}(\bm{B})$.
 Finally, the last term of (\ref{fropt1}) can be re-expressed as
 \begin{align}
&-2\mathfrak{Re}\{ \bm{\nu}_{\zeta}^H\bm{R\Phi u}_{\zeta}\}=-2\mathfrak{Re}\{\mathrm{Tr}(\bm{u}_{\zeta} \bm{\nu}_{\zeta}^H\bm{R\Phi})\}\\
&\quad\quad\  \quad=-2\mathfrak{Re}\Big\{\mathrm{vec}^H(\bm{R}^H\bm{\nu}_{\zeta}\bm{u}_{\zeta}^H)\Big(\sum_{g=1}^G \bm{Q}_g \bm{D}_g\bm{\phi}_g\Big)\Big\},\!\!\label{fropt2vec}
\end{align}
which holds due to $\mathrm{Tr} (\bm{A}^H \bm{B})=\mathrm{vec}^H(\bm{A})\mathrm{vec}(\bm{B})$. 

Consequently, $f_{\zeta}(\bm{\Phi},\bm{\nu}_{\zeta})$ can be re-expressed as
\begin{align}
	&\!\!\!\!f_{\zeta}(\{\bm{\phi}_g\}_{g=1}^G, \bm{\nu}_{\zeta})	\!=\!\Big(\sum_{g=1}^G \bm{Q}_g \bm{D}_g\bm{\phi}_g\Big)^H\bm{V}_{\zeta}\Big(\sum_{g=1}^G \bm{Q}_g \bm{D}_g\bm{\phi}_g\Big)\nonumber\\
	&\!\!\!\!\!\!-\!\suo 2\mathfrak{Re}\suo\Big\{\suo\bm{q}_{\zeta}^H\suo\Big(\suo\sum_{g=1}^G\suo \bm{Q}_g \bm{D}_g\bm{\phi}_g\suo\Big)\suo\Big\}\!\suo+\suo\!\sum\limits_{k=1}^K \suo P_{k} \suo\vert\bm{\nu}_{\zeta}^H \suo \bm{h}_{\mathrm{d},k}\suo\vert^2\!+\!\sigma^2\suo\Vert\bm{\nu}_{\zeta}\suo\Vert^2,\!\!\!
\end{align}
where $\bm{q}_{\zeta}^H\triangleq\mathrm{vec}^H\big(\bm{R}^H\bm{\nu}_{\zeta}(\bm{u}_{\zeta}^H-\sum_{k=1}^K P_{k} \bm{\nu}_{\zeta}^H \bm{h}_{\mathrm{d},k}\bm{h}^H_{\mathrm{r},k})\big)$.
Similarly, $f_{k}(\bm{\Phi},\bm{\nu}_{k})$ can be re-expressed as
\begin{align}
	&\!\!\!\!f_k\suo(\suo\{\bm{\phi}_g\suo\}_{g=1}^G\suo,\suo\bm{\nu}_k\suo)\suo\!=\suo\!\Big(\suo\!\sum_{g=1}^G \!\bm{Q}_g \bm{D}_g\bm{\phi}_g\!\suo\Big)^H\!\!\bm{J}_k\suo\Big(\suo\!\sum_{g=1}^G \!\bm{Q}_g \bm{D}_g\bm{\phi}_g\!\suo\Big)\!\!+\!\sigma^2\suo\Vert\bm{\nu}_k\suo\Vert^2\nonumber\\
	&\!\!\!-\!2\mathfrak{Re}\Big\{\!\bm{r}_k^H\!\Big(\!\sum_{g=1}^G \!\bm{Q}_g \bm{D}_g\bm{\phi}_g\!\Big)\!\!+\!\bm{\nu}_k^H \!\bm{h}_{\mathrm{d},k}\!\Big\}
	\!\suo+\!\!\sum\limits_{\substack{k'=1\\k'\neq k}}^{K}\!\! P_{k'}\suo\vert\bm{\nu}_k^H\!\bm{h}_{\mathrm{d},k'}\!\vert^2\suo,\!\!\!\!
\end{align}
where $\bm{J}_k\triangleq\sum_{k'=1,k'\neq k}^{K} P_{k'}  \bm{J}_{k,k'}  +  P_0  \bm{G}^T \otimes (\bm{R}^H \bm{\nu}_k \bm{\nu}_k^H \bm{R})$, $\bm{J}_{k,k'}\triangleq(\bm{h}_{\mathrm{r},k'}^*\bm{h}_{\mathrm{r},k'}^T) \otimes (\bm{R}^H \bm{\nu}_k \bm{\nu}_k^H \bm{R})$, and $\bm{r}_k^H \triangleq \mathrm{vec}^H\big(\bm{R}^H\bm{\nu}_k(\bm{h}_{\mathrm{r},k}^H-\sum_{k'=1,k'\neq k}^K P_{k'} \bm{\nu}_{k}^H \bm{h}_{\mathrm{d},k'}\bm{h}^H_{\mathrm{r},k'})\big),\ k=1,...,K$. 
Note that $\bm{J}_k$'s are also PSD matrices. In addition, the objective function of Problem (P4-I-eqv) can be re-expressed as $-\alpha+\sum_{g=1}^G\mathfrak{Re}\big\{\mathrm{Tr}\big(\bm{\Lambda}_g^H(\bm{\Phi}_g-\bm{\Psi}_g)\big)\big\}+\sum_{g=1}^G\frac{1}{2\rho}\Vert \bm{\Phi}_g-\bm{\Psi}_g \Vert^2_\mathrm{F}=-\alpha+\sum_{g=1}^G\mathfrak{Re}\big\{\mathrm{vec}^H(\bm{\Lambda}_g)\big(\bm{D}_g\bm{\phi}_g-\mathrm{vec}(\bm{\Psi}_g)\big)\big\}+\sum_{g=1}^G\frac{1}{2\rho}\Vert \bm{D}_g\bm{\phi}_g-\mathrm{vec}(\bm{\Psi}_g) \Vert^2$ due to $\mathrm{Tr} (\bm{A}^H \bm{B})=\mathrm{vec}^H(\bm{A})\mathrm{vec}(\bm{B})$.

Therefore, with given $(\{\bm{\nu}_{\zeta}\}_{\zeta=1}^{R}, \{\bm{\nu}_k\}_{k=1}^K)$, $(\alpha,\bm{\Phi})$ can be optimized by solving the following problem on $(\alpha,\{\bm{\phi}_g\}_{g=1}^G)$ and obtaining the optimal $\bm{\Phi}$ via (\ref{recover}):
%\begin{align}
%	\mbox{(P4-I-eqv')}  \min_{\alpha,\{\bm{\phi}_g\}_{g=1}^G}& -\alpha+\sum\limits_{g=1}^G\mathfrak{Re}\big\{\mathrm{vec}^H(\bm{\Lambda}_g)\big(\bm{D}_g\bm{\phi}_g-\mathrm{vec}(\bm{\Psi}_g)\big)\big\}+\sum\limits_{g=1}^G\frac{1}{2\rho}\Vert \bm{D}_g\bm{\phi}_g-\mathrm{vec}(\bm{\Psi}_g) \Vert^2
%	\label{P8obj}\\
%	\mathrm{s.t.}\ \ &\  \Big(\sum_{g=1}^G \bm{Q}_g \bm{D}_g\bm{\phi}_g\Big)^H\bm{J}_k\Big(\sum_{g=1}^G \bm{Q}_g \bm{D}_g\bm{\phi}_g\Big)\!-\!2\mathfrak{Re}\Big\{\bm{r}_k^H\Big(\sum_{g=1}^G \bm{Q}_g \bm{D}_g\bm{\phi}_g\Big)+\bm{\nu}_k^H \bm{h}_{\mathrm{d},k}\!\Big\}\nonumber\\
%	&\ +\sum\limits_{\substack{k'=1\\k'\neq k}}^{K} P_{k'}\vert\bm{\nu}_k^H\bm{h}_{\mathrm{d},k'}\vert^2+\sigma^2\Vert\bm{\nu}_k\Vert^2+\frac{\alpha}{P_{k}}\leq 0,\  k=1,...,K\label{SINRcons8}\\
%	&\ \sum\limits_{\zeta=1}^{R} \kappa_{\zeta}\Big\{\Big(\sum_{g=1}^G \bm{Q}_g \bm{D}_g\bm{\phi}_g\Big)^H\bm{V}_{\zeta}\Big(\sum_{g=1}^G \bm{Q}_g \bm{D}_g\bm{\phi}_g\Big)-2\mathfrak{Re}\Big\{\bm{q}_{\zeta}^H\Big(\sum_{g=1}^G \bm{Q}_g \bm{D}_g\bm{\phi}_g\Big)\Big\}\nonumber\\
%	&\ +\sum\limits_{k=1}^K P_{k} \vert\bm{\nu}_{\zeta}^H \bm{h}_{\mathrm{d},k}\vert^2+\sigma^2\Vert\bm{\nu}_{\zeta}\Vert^2\Big\}+\Gamma\leq 0\label{PCRBcons8}.
%\end{align}
\begin{align}
	&\hspace{-1.8cm}\mbox{(P4-I-eqv')}\nonumber\\
\hspace{-2cm}\min_{\alpha,\{\bm{\phi}_g\}_{g=1}^G}\hspace{.3cm}&\hspace{-.3cm} -\alpha+\sum\limits_{g=1}^G\mathfrak{Re}\big\{\mathrm{vec}^H(\bm{\Lambda}_g)\big(\bm{D}_g\bm{\phi}_g-\mathrm{vec}(\bm{\Psi}_g)\big)\big\}\hspace{-.7cm}\nonumber\\
	&\hspace{.685cm}+\sum\limits_{g=1}^G\frac{1}{2\rho}\Vert \bm{D}_g\bm{\phi}_g-\mathrm{vec}(\bm{\Psi}_g) \Vert^2
	\label{P8obj}\\
	\mathrm{s.t.}\quad&\Big(\suo\sum_{g=1}^G \!\bm{Q}_g \bm{D}_g\bm{\phi}_g\suo\Big)^H\!\bm{J}_k\suo\Big(\suo\sum_{g=1}^G \bm{Q}_g \bm{D}_g\bm{\phi}_g\suo\Big)+\sigma^2\Vert\bm{\nu}_k\suo\Vert^2\!\!\!\!\!\!\!\!\!\!\!\!\!\!\!\!\nonumber\\
	&\ -2\mathfrak{Re}\Big\{\suo\bm{r}_k^H\suo\Big(\suo\sum_{g=1}^G \bm{Q}_g \bm{D}_g\bm{\phi}_g\suo\Big)+\bm{\nu}_k^H \bm{h}_{\mathrm{d},k}\suo\Big\}\nonumber\\
	&\ +\!\suo\sum\limits_{\substack{k'=1\\k'\neq k}}^{K}\!\! P_{k'}\suo\vert\bm{\nu}_k^H\suo\bm{h}_{\mathrm{d},k'}\suo\vert^2\!+\!\suo\frac{\alpha}{P_{k}}\!\leq\! 0,\ k=1,...,K\!\!\!\!\!\!\!\!\!\label{SINRcons8}\\
	&\sum\limits_{\zeta=1}^{R} \kappa_{\zeta}\Big\{\Big(\sum_{g=1}^G \bm{Q}_g \bm{D}_g\bm{\phi}_g\big)^{H}\bm{V}_{\zeta}\Big(\sum_{g=1}^G \bm{Q}_g \bm{D}_g\bm{\phi}_g\Big)\hspace{-.7cm}\nonumber\\
	&\ -2\mathfrak{Re}\Big\{\bm{q}_{\zeta}^H\Big(\sum_{g=1}^G \bm{Q}_g \bm{D}_g\bm{\phi}_g\Big)\Big\}+\sigma^2\Vert\bm{\nu}_{\zeta}\Vert^2\hspace{-.5cm}\nonumber\\
	&\ +\sum\limits_{k=1}^K P_{k} \vert\bm{\nu}_{\zeta}^H \bm{h}_{\mathrm{d},k}\vert^2\Big\}+\Gamma\leq 0\label{PCRBcons8}.
\end{align}

By noting $\bm{J}_k$'s and $\bm{V}_{\zeta}$ are PSD matrices, (P4-I-eqv') can be shown to be a convex second-order cone program (SOCP), for which the optimal solution can be obtained via the interior-point method \cite{complexity} or existing software such as CVX \cite{cvx}. 

\subsubsection{Overall Algorithm for Problem (P4-I-eqv)}
To summarize, the optimal solutions to $(\alpha,\bm{\Phi})$ and $(\{\bm{\nu}_{\zeta}\}_{\zeta=1}^{R},\{\bm{\nu}_k\}_{k=1}^K)$ can be iteratively obtained with the other being fixed at each time. Since the objective value of Problem (P4-I-eqv) is bounded below, monotonic convergence is guaranteed for this alternating optimization algorithm.

\vspace{-.3cm}
\subsection{(Inner Loop) Optimization of Auxiliary Variables $\{\bm{\Psi}_g\}_{g=1}^G$ with Given $\bm{\Phi}$}
Note that $\mathfrak{Re}\big\{\mathrm{Tr}\big(\bm{\Lambda}_g^H(\bm{\Phi}_g-\bm{\Psi}_g)\big)\big\}+\frac{1}{2\rho}\Vert \bm{\Phi}_g-\bm{\Psi}_g \Vert^2_\mathrm{F}=\frac{1}{2\rho}\big(\Vert \bm{\Psi}_g-(\bm{\Phi}_g+\rho \bm{\Lambda}_g)\Vert_\mathrm{F}^2 -\rho^2 \mathrm{Tr}(\bm{\Lambda}_g^H \bm{\Lambda}_g)  \big)$ holds.  Thus, with given $\bm{\Phi}$, the sub-problem for optimizing $\bm{\Psi}_g$ is given by
\begin{align}
	\text{(P4-II)}\quad \min_{\bm{\Psi}_g}& \quad \Vert \bm{\Psi}_g-(\bm{\Phi}_g+\rho \bm{\Lambda}_g)\Vert_\mathrm{F}^2\\
	\mathrm{s.t.}&\quad \bm{\Psi}_g^H\bm{\Psi}_g=\bm{I}_{M_g}.
\end{align}
Denote $\bm{\Phi}_g+\rho \bm{\Lambda}_g= \tilde{\bm{U}}_g\tilde{\bm{S}}_g\tilde{\bm{V}}_g^H$ as the singular value decomposition (SVD) of $\bm{\Phi}_g+\rho \bm{\Lambda}_g$. Note that the objective function of (P4-II) is convex, while the only constraint is a unitary constraint. According to \cite{unitary}, the optimal solution to (P4-II) can be derived in closed form as
\begin{gather}
	\bm{\Psi}_g^\star = \tilde{\bm{U}}_g\tilde{\bm{V}}^H_g.\label{optPsi}
\end{gather}

\vspace{-.3cm}
\subsection{(Outer Loop) Update of Dual Variables $\{\bm{\Lambda}_g\}_{g=1}^G$ or Penalty Parameter $\rho$}
In each outer iteration, we only update either the dual variables $\{\bm{\Lambda}_g\}_{g=1}^G$ or the penalty parameter $\rho$
based on a predefined constraint violation level $\epsilon$ \cite{penalty1}. If the equality constraints in (\ref{equal}) are approximately satisfied with a violation lower than $\epsilon$, i.e., $\Vert\bm{\Phi}_g-\bm{\Psi}_g\Vert_\infty\leq \epsilon,\ g=1,...,G$, each dual variable $\bm{\Lambda}_g$ is updated as
\begin{gather}
	\bm{\Lambda}_g\leftarrow \bm{\Lambda}_g + \frac{1}{\rho}(\bm{\Phi}_g-\bm{\Psi}_g).\label{lambdaupdate}
\end{gather}
Otherwise, $\rho$ is decreased as follows to enhance the penalty term to push the equality constraint towards being satisfied:
\begin{gather}
	\rho \leftarrow \delta\rho\label{rhoupdate},
\end{gather}
where $0<\delta<1$ is the shrinkage factor.

%\begin{table}[t]
%	\centering
%	\caption{Complexity Analysis for Proposed PDD-based Algorithm for Problem (P1)}
%	%	\vspace{-.3cm}
%	
%	\label{complexity}
%	\begin{tabular}{|c|c|c|c|}
%		\cline{1-4}
%		&Optimizing $\{\bm{\nu}_{\zeta}\}_{\zeta=1}^{R}, \{\bm{\nu}_k\}_{k=1}^K$&Optimizing $(\alpha,\bm{\Phi})$ & Optimizing $\{\bm{\Psi}_g\}_{g=1}^G$\\
%		\hline
%		Complexity & $\mathcal{O}(N^3)$ & $\mathcal{O}\big(K^{1.5}(\sum_{g=1}^G M_g^2)^3\big)$ &$\sum_{g=1}^G M_g^3$ \\
%		\hline
%		Overall Complexity for Problem (P1) &\multicolumn{3}{c|}{$\mathcal{O}\Big(L_\mathrm{P}L_\mathrm{O}L_\mathrm{I}L_\mathrm{P4}\big(N^3+K^{1.5}(\sum_{g=1}^G M_g^2)^3\big)\Big)$}  \\
%		\hline
%	\end{tabular}
%	%	\vspace{-.2cm}
%\end{table}
%\begin{table}[t]
%	\centering
%	\caption{Complexity Analysis for Proposed PDD-based Algorithm for Problem (P1)}
%	%	\vspace{-.3cm}
%	
%	\label{complexity}
%	\begin{tabular}{|c|c|c|c|}
%			\cline{1-4}
%			&Optimizing $\{\bm{\nu}_{\zeta}\}_{\zeta=1}^{R}, \{\bm{\nu}_k\}_{k=1}^K$&Optimizing $(\alpha,\bm{\Phi})$ & Optimizing $\{\bm{\Psi}_g\}_{g=1}^G$\\
%			\hline
%			Complexity & $\mathcal{O}(N^3)$ & $\mathcal{O}\big(K^{1.5}(\sum_{g=1}^G M_g^2)^3\big)$ &$\sum_{g=1}^G M_g^3$ \\
%			\hline
%			Overall Complexity for Problem (P1) &\multicolumn{3}{c|}{$\mathcal{O}\Big(L_\mathrm{P}L_\mathrm{O}L_\mathrm{I}L_\mathrm{P4}\big(N^3+K^{1.5}(\sum_{g=1}^G M_g^2)^3\big)\Big)$}  \\
%			\hline
%		\end{tabular}
%		\vspace{-.2cm}
%\end{table}

\begin{table}[t]
	\centering
	\caption{Complexity Analysis for Proposed PDD-based Algorithm for Problem (P1)}
		\vspace{-.2cm}
	
	\label{complexity}
	\begin{tabular}{|c|c|}
		\hline
		Optimization Part&Complexity\\
		\hline
		 $\!\{\bm{\nu}_{\zeta}\}_{\zeta=1}^{R}, \{\bm{\nu}_k\}_{k=1}^K\!$ & $\mathcal{O}(N^3)$\\
		 \hline
		 $(\alpha,\bm{\Phi})$ & $\mathcal{O}(K^{1.5}(\sum_{g=1}^G M_g^2)^3)$\\
		 \hline
		 $\{\bm{\Psi}_g\}_{g=1}^G$ & $\sum_{g=1}^G M_g^3$\\
		 \hline
		Problem (P1) & $\!\mathcal{O}\big(L_\mathrm{P}L_\mathrm{O}L_\mathrm{I}L_\mathrm{P4}(N^3+K^{1.5}(\sum_{g=1}^G M_g^2)^3)\big)\!$\\
		\hline
	\end{tabular}
	\vspace{-.5cm}
\end{table}

\vspace{-.3cm}
\subsection{Overall PDD-based Algorithm for Problem (P1)}
To summarize, the overall PDD-based algorithm for (P4) consists of two loops. In the outer loop, the dual variables in $\{\bm{\Lambda}_g\}_{g=1}^G$ or the penalty parameter $\rho$ is updated based on the constraint violation level of (\ref{equal}) according to \eqref{lambdaupdate} and \eqref{rhoupdate}. In the inner loop, the optimal solutions to $(\alpha,\bm{\Phi})$, auxiliary vectors $(\{\bm{\nu}_{\zeta}\}_{\zeta=1}^{R}, \{\bm{\nu}_k\}_{k=1}^K)$, and auxiliary matrices $\{\bm{\Psi}_g\}_{g=1}^G$ are iteratively obtained by solving the convex problem (P4-I-eqv') and recovering $\bm{\Phi}$ via (\ref{recover}) or according to the closed-form expressions in (\ref{lambdar}), (\ref{lambdak}), and (\ref{optPsi}), respectively. Note that as the optimal solution to each sub-problem is obtained in each step, convergence can be achieved for the proposed PDD-based algorithm, which will be verified in Section VII. By updating the penalty parameter $\rho$, the constraint violation level can be controlled under the given threshold $\epsilon$. Moreover, to further enhance the performance, the above procedure can be repeated for $L_{\mathrm{P}}>1$ times with different initialization, among which the best solution can be selected, while we will present one initialization method in Section \ref{PCRB Minimization}. Finally, the obtained solution of $\bm{\Phi}$ for (P4) automatically serves as a high-quality suboptimal solution to (P1). Let $L_\mathrm{O}$ and $L_\mathrm{I}$ denote the numbers of outer and inner loops in the proposed algorithm. Let $L_\mathrm{P4}$ denote the iteration numbers for solving (P4-I-eqv). The overall complexity of the proposed algorithm is analyzed in Table \ref{complexity}, which is observed to be polynomial over key system parameters $N$, $K$, and $M_g$'s.

\vspace{-.3cm}
\subsection{Extension to BD-IRS Aided Uplink Sensing for PCRB Minimization}\label{PCRB Minimization}
Finally, we note that the proposed algorithm can be extended to deal with the PCRB minimization problem in a BD-IRS aided uplink sensing system, where no communication user is present. The problem is formulated as
\begin{align}
	\!	\!\!\!\mbox{(P5)}\    \min_{\bm{\Phi}}\ &\frac{1}{2P_0L\!\sum\limits_{\zeta=1}^{R}\! \kappa_{\zeta}(\bm{R}\bm{\Phi}\bm{u}_{\zeta})^H \bm{\Sigma}_0^{-1}\suo(\bm{\Phi}) (\bm{R}\bm{\Phi}\bm{u}_{\zeta})\!+\!F_\mathrm{P}}\!\!\\
	\mathrm{s.t.}\ & \bm{\Phi}=\mathrm{blkdiag}\{\bm{\Phi}_1,...,\bm{\Phi}_G\}\\
	&\bm{\Phi}_g^H\bm{\Phi}_g = \bm{I}_{M_g},\ \quad g=1,...,G\\
	&\bm{\Phi}_g=\bm{\Phi}_g^T,\hspace{0.78cm}\quad g=1,...,G.
\end{align}
Notice that the objective PCRB function of (P5) appears in the constraints of (P2). Thus, by introducing auxiliary variables $\bm{\Psi}_g$'s and $\bm{\nu}_\zeta$'s, we can formulate the AL problem for (P5) under the PDD framework and obtain a high-quality solution to (P5) in a similar manner, for which the details are omitted due to limited space. Note that the solution to (P5) can serve as an initial point for the proposed algorithm for (P1), as it aims to achieve the minimum PCRB value and consequently satisfy the PCRB constraint in (\ref{PCRBcons1}).

\section{BD-IRS Aided Uplink ISAC under TDMA}\label{sec6.5}
In the previous sections, we studied an SDMA scheme where the interference between sensing and communication is managed via BD-IRS reflection and BS receive beamforming optimization. However, the new mutual interference between sensing and communication is more severe than the one-way interference from sensing to communication in mono-static downlink ISAC. Particularly, as BD-IRS has a constrained circuit architecture, its effectiveness in simultaneously two conflicting goals in sensing and communication is generally limited. This thus motivates us to explore whether other strictly orthogonal multiple access schemes can achieve enhanced performance via removing such mutual interference.

\begin{figure}[tbp]
	\centering
	\includegraphics[width=\linewidth]{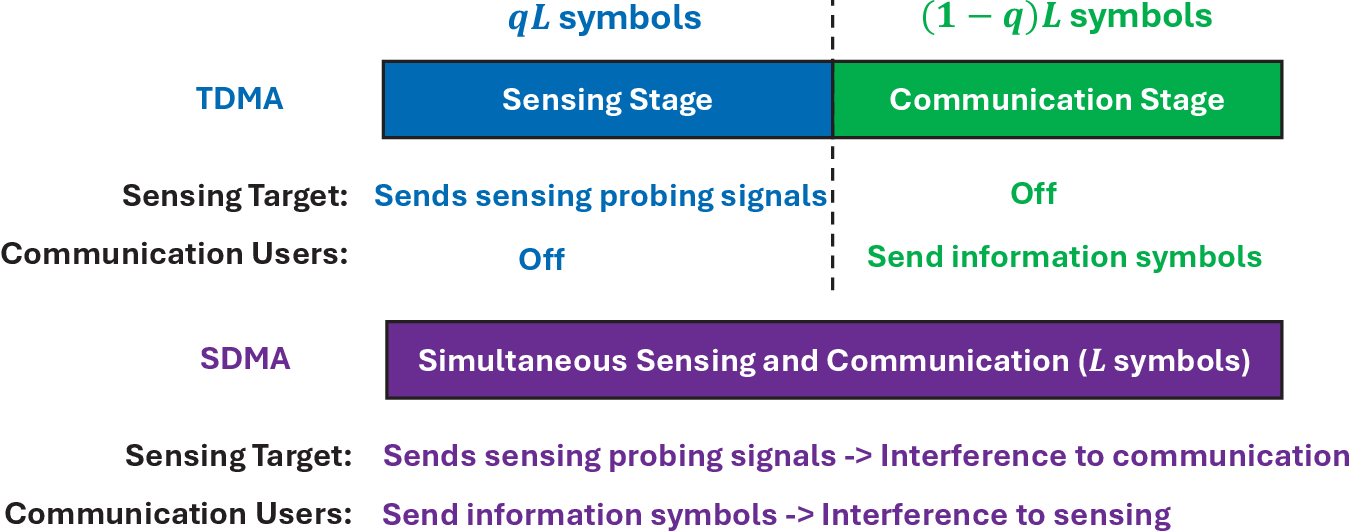}
	\vspace{-5mm}
	\caption{Comparison between the SDMA scheme and TDMA scheme.}
	\label{TDMAillustration}
	\vspace{-.5cm}
\end{figure}

In this section, we propose a TDMA scheme for uplink BD-IRS aided ISAC illustrated in Fig. \ref{TDMAillustration}, where the sensing target sends uplink signals for sensing during $qL$ symbol intervals (i.e., the \emph{sensing stage}), where $q\in [0,1]$ denotes the fraction of time for sensing, while the $K$ communication users send uplink signals during the other $(1-q)L$ symbol intervals (i.e., the \emph{communication stage}).
 In this case, two different BD-IRS reflection matrices denoted by $\bm{\Phi}_{\mathrm{S}}$ and $\bm{\Phi}_{\mathrm{C}}$ can be separately designed to optimize the sensing PCRB and minimum communication user rate, and implemented in the sensing stage and communication stage, respectively.

During the sensing stage, the interference from communication users is not present, and the interference-plus-noise covariance matrix reduces to the noise covariance matrix $\bm{\Sigma}_0(\bm{\Phi}_{\mathrm{S}})=\sigma^2\bm{I}_N$. Thus, the PCRB expression is given by $
	\mathrm{PCRB}^{\mathrm{TDMA}}(\bm{\Phi}_\mathrm{S}) = \frac{1}{\frac{2P_0qL}{ \sigma^2}\sum_{\zeta=1}^{R} \kappa_{\zeta}(\bm{R}\bm{\Phi}_{\mathrm{S}}\bm{u}_{\zeta})^H  \bm (\bm{R}\bm{\Phi}_{\mathrm{S}}\bm{u}_{\zeta})+F_\mathrm{P}}$. Then, $\bm{\Phi}_{\mathrm{S}}$ can be optimized to minimize $\mathrm{PCRB}^{\mathrm{TDMA}}(\bm{\Phi}_\mathrm{S})$ using a similar method as in Section \ref{PCRB Minimization}.
	
	During the communication stage, the SINR given below is no longer a random variable as the sensing-generated interference due to the randomly located target is not present:
\begin{align}
	\gamma_k^{\mathrm{C}}(\bm{w}_k,\bm{\Phi}_{\mathrm{C}})&=\frac{P_{k} \vert \bm{w}_k^H \bm{h}_k(\bm{\Phi}_{\mathrm{C}})\vert^2}{\sum\limits_{\substack{k'=1\\k' \neq k}}^K P_{k'} \vert \bm{w}_k^H \bm{h}_{k'}(\bm{\Phi}_{\mathrm{C}}) \vert^2 + \sigma^2\Vert \bm{w}_k\Vert^2}\\
	&=\frac{P_{k} \vert \bm{w}_k^H \bm{h}_k(\bm{\Phi}_{\mathrm{C}})\vert^2}{\bm{w}_k^H \bm{\Sigma}_k^{\mathrm{C}}(\bm{\Phi}_{\mathrm{C}}) \bm{w}_k},
\end{align}
where $\bm{\Sigma}_k^\mathrm{C}(\bm{\Phi}_{\mathrm{C}}) =  \sum_{k'=1,k' \neq k}^K P_{k'} \bm{h}_{k'}(\bm{\Phi}_{\mathrm{C}})\bm{h}_{k'}^H(\bm{\Phi}_{\mathrm{C}})+\sigma^2\bm{I}_N,\ k=1,...,K$. Given optimal $\bm{w}_k^\star=\frac{{\bm{\Sigma}_k^\mathrm{C}}^{-1}(\bm{\Phi}_{\mathrm{C}})\bm{h}_k(\bm{\Phi}_{\mathrm{C}})}{\Vert{\bm{\Sigma}_k^\mathrm{C}}^{-1}(\bm{\Phi}_{\mathrm{C}})\bm{h}_k(\bm{\Phi}_{\mathrm{C}})\Vert}$, the effective achievable rate is expressed as follows in bps/Hz:
\begin{align}
	R_k^\mathrm{TDMA}(\bm{\Phi}_{\mathrm{C}})\!=\!(1\!-\!q)\suo\log_2\!\big(\suo 1\suo\!+\!\suo P_{k} \bm{h}_k^H\!(\suo\bm{\Phi}_{\mathrm{C}}\suo){\bm{\Sigma}_k^\mathrm{C}}^{-1}\! (\suo\bm{\Phi}_\mathrm{C}\suo)\bm{h}_k\suo(\suo\bm{\Phi}_{\mathrm{C}}\suo)\big).
\end{align}
Then, $\bm{\Phi}_{\mathrm{C}}$ can be designed to maximize the minimum rate among $R_k^\mathrm{TDMA}(\bm{\Phi}_{\mathrm{C}})$'s, which can be done via a similar PDD-based algorithm as in Section V.

Note that increasing the time allocated to sensing (i.e., increasing $q$) will reduce the PCRB, but also leads to smaller communication rates. The optimal TDMA time allocation can be obtained via solving the following problem:
\begin{align}
	\mbox{(P6)}\  \max_{ q\in [0,1]}\  \underset{k=1,...,K}{\min}\, &(1-q) \log_2 \big(  1+\nonumber\\
	&\  P_{k} \bm{h}_k^H ( \bm{\Phi}_{\mathrm{C}} ){\bm{\Sigma}_k^\mathrm{C}}^{-1} ( \bm{\Phi}_\mathrm{C} )\bm{h}_k ( \bm{\Phi}_{\mathrm{C}} )\big)\!\!\\
	\mathrm{s.t.}\ \ \,& \frac{1}{\frac{2P_0qL}{ \sigma^2}\!\sum\limits_{\zeta=1}^{R}\! \kappa_{\zeta}(\suo\bm{R}\bm{\Phi}_{\mathrm{S}}\bm{u}_{\zeta}\suo)^H \suo\bm (\suo\bm{R}\bm{\Phi}_{\mathrm{S}}\bm{u}_{\zeta}\suo)\!+\!F_\mathrm{P}}\nonumber\\
	&\hspace{3.25cm}\leq \Gamma_{\mathrm{PCRB}}.\label{TDMA-PCRBcons}
\end{align}
Note that both the objective function to be maximized and the left-hand-side of the constraint in \eqref{TDMA-PCRBcons} are non-increasing functions of $q$. Thus, (P6) is feasible if and only if $$\frac{1}{\frac{2P_0L}{ \sigma^2}\sum\limits_{\zeta=1}^{R} \kappa_{\zeta}(\bm{R}\bm{\Phi}_{\mathrm{S}}\bm{u}_{\zeta})^H \bm (\bm{R}\bm{\Phi}_{\mathrm{S}}\bm{u}_{\zeta})+F_\mathrm{P}}\leq \Gamma_{\mathrm{PCRB}}$$ holds, and its optimal solution is given by\footnote{The obtained number of symbol intervals can be rounded up to the nearest integer if needed.}
\begin{align}
	q^\star=\max\bigg\{\frac{\frac{1}{\Gamma_{\mathrm{PCRB}}}-F_\mathrm{P}}{\frac{2P_0L}{\sigma^2}\sum_{\zeta=1}^{R} \kappa_{\zeta}(\bm{R}\bm{\Phi}_{\mathrm{S}}\bm{u}_{\zeta})^H \bm (\bm{R}\bm{\Phi}_{\mathrm{S}}\bm{u}_{\zeta})},0\bigg\}.
\end{align}

Note that although a pre-log factor $1-q$ is applied in the achievable rate, the removal of interference in both ${\bm{\Sigma}_k^\mathrm{C}}(\bm{\Phi}_{\mathrm{C}})$'s and $\bm{\Sigma}_0(\bm{\Phi}_{\mathrm{S}})$ as well as the separate tailored design of BD-IRS reflection matrix in the two stages also bring new benefits. In the next section, we will compare the performance of TDMA versus SDMA to draw more useful insights.

\section{Numerical Results}\label{sec6}

\begin{figure}[t]
	\centering
	\includegraphics[width=.7\linewidth]{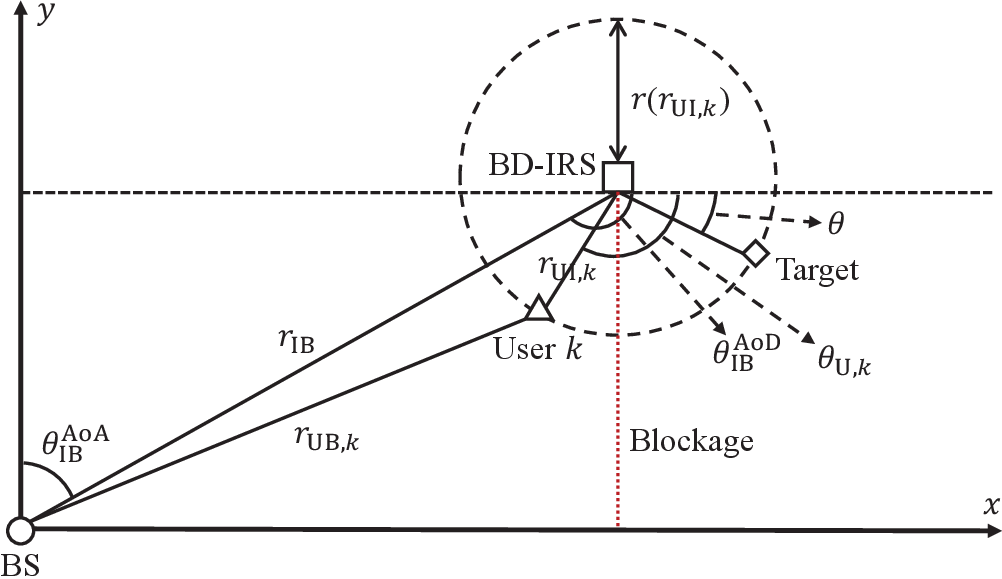}
	\vspace{-2mm}
	\caption{Topology of the BD-IRS aided uplink ISAC system.}\label{simulation}
	\vspace{-.3cm}
\end{figure}

In this section, we provide numerical results to evaluate the effectiveness of our proposed BD-IRS reflection matrix design. Under a two-dimensional Cartesian coordinate system, the topology of the system is illustrated in Fig. \ref{simulation}, where the distance between the BD-IRS and the BS is set as $r_{\mathrm{IB}}=200$ m, the distance between the possible locations of the target as well as the communication users and the BD-IRS is set as $r=r_{\mathrm{UI},k}=10\text{ m},\ k=1,...,K$. The BS is equipped with a uniform linear array (ULA) vertical to the $x$-axis, while the BD-IRS lies on the $x-z$ plane. The spacings between adjacent BS antennas and BD-IRS elements are set as $\Delta =\frac{\lambda}{2}$. Let $\theta_{\mathrm{U},k}$ denote the angle-of-arrival (AoA) from each $k$-th user to the BD-IRS. Let  $\theta^\mathrm{AoA}_{\mathrm{IB}}=\frac{\pi}{4}$ and $\theta^\mathrm{AoD}_{\mathrm{IB}}=\theta^\mathrm{AoA}_{\mathrm{IB}}+\frac{\pi}{2}$ denote the AoA and angle-of-departure (AoD) of the link from the BD-IRS to the BS, respectively.  Then, the distance between each $k$-th user and the BS can be calculated as $r_{\mathrm{UB},k}=\sqrt{r_{\mathrm{IB}}^2+ r_{\mathrm{UI},k}^2-2r_{\mathrm{IB}}r_{\mathrm{UI},k}\cos (\theta^\mathrm{AoD}_{\mathrm{IB}}-\theta_{\mathrm{U},k})}$. We consider a Rician fading model for the channel from the BD-IRS to the BS with $\bm{R}=\beta_{\mathrm{IB}}\sqrt{\frac{1}{\chi+1}}(\sqrt{\chi}\bm{H}_\mathrm{LoS}+\bm{H}_\mathrm{NLoS})$, where $\chi=-8$ dB is the Rician factor, $\beta_{\mathrm{IB}}$ denotes the path gain given by $\beta_{\mathrm{IB}}=\frac{\beta_0}{{r_\mathrm{IB}}}$ with $\beta_0=-33$ dB, $\bm{H}_{\mathrm{LoS}}$ represents the LoS component, and $\bm{H}_{\mathrm{NLoS}}$ represents the non-LoS (NLoS) component following Rayleigh fading, respectively. The LoS component can be expressed as  $\bm{H}_{\mathrm{LoS}}=\bm{a}(\theta^\mathrm{AoA}_{\mathrm{IB}})\bm{b}^H(\theta^\mathrm{AoD}_{\mathrm{IB}})$, where $a_n(\theta^\mathrm{AoA}_{\mathrm{IB}})=e^{j\pi(n-1)\cos \theta^\mathrm{AoA}_{\mathrm{IB}}},\ n=1,...,N$ and $b_m(\theta^\mathrm{AoD}_{\mathrm{IB}})=e^{j\pi (m-1)\mathrm{mod}M_x\cos \theta^\mathrm{AoD}_{\mathrm{IB}} },\ m=1,...,M$ are steering vectors at the BS and the BD-IRS, respectively. We consider a Rayleigh fading model for the channels from the users to the BS with path loss exponent $3.5$, and consider an LoS model for the channels from the users to the BD-IRS with ${h_{\mathrm{r},k}}_m(\theta_{\mathrm{U},k}) =\frac{\beta_0}{{r_{\mathrm{UI},k}}} e^{j\pi (m-1)\mathrm{mod}M_x\cos \theta_{\mathrm{U},k} },\ m=1,...,M$.

We consider $K=2$ communication users, $N=16$ receive antennas at the BS, $P_0=P_{1}=P_{2}=10$ dBm, $\sigma^2=-95$ dBm, $L=25$, and $M_x=M_z=4$ if not specified otherwise. The azimuth angles of users are set as $\theta_{\mathrm{U},1}=\frac{5\pi}{9}$ and $\theta_{\mathrm{U},2}=\frac{7\pi}{9}$. The PDF of $\theta$ is assumed to follow the Gaussian mixture model in Remark 1 with $p_1=0.31,\  \theta_1=\frac{5\pi}{18},\ \sigma^2_1=10^{-3}$, $p_2=0.43, \ \theta_2=\frac{11\pi}{36},\ \sigma^2_2=10^{-3}$, and $p_3=0.26, \ \theta_3=\frac{\pi}{3},\ \sigma^2_3=10^{-3}$. We set  $\Gamma_{\mathrm{PCRB}}=5\times 10^{-4}$ and consider the fully-connected case unless otherwise specified.

For comparison with the proposed design, we consider the following two benchmark schemes:
\begin{itemize}
	\item \textbf{Benchmark Scheme 1: Isotropic reflection.} We set $\bm{\Phi}=\bm{I}_M$ which corresponds to isotropic reflection and can also be realized using conventional diagonal IRS.
	\item \textbf{Benchmark Scheme 2: Random reflection.} We randomly generate $100$ BD-IRS reflection matrices under the fully-connected architecture, and select the best one among them.
\end{itemize}

\vspace{-.3cm}

\subsection{Convergence Behavior of Proposed PDD-based Algorithms}
\begin{figure*}[t]
	\centering
	\setcounter{figure}{4}
	\addtolength{\subfigcapskip}{-1mm}
	\subfigure[PCRB versus $M_x$.]{
		\includegraphics[width=.315\linewidth]{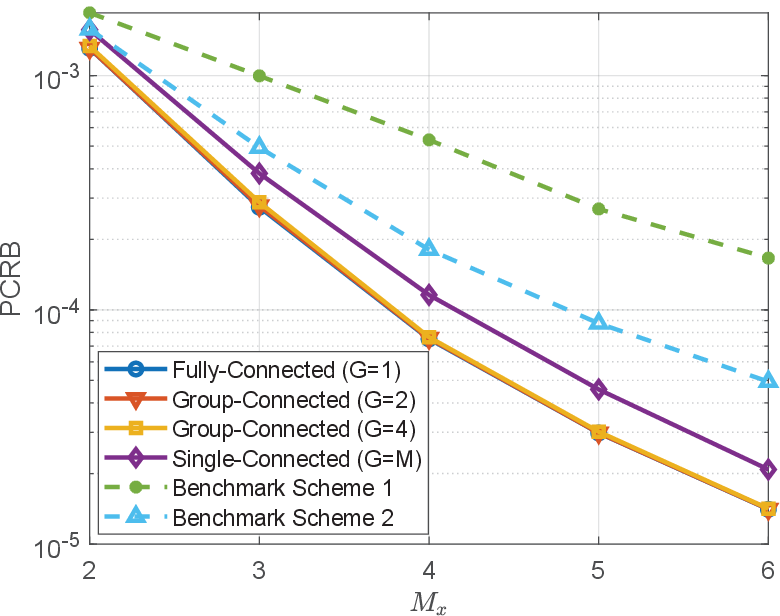}	\label{sensing_Mx}}
	\subfigure[PCRB versus $M_z$.]{
		\includegraphics[width=.315\linewidth]{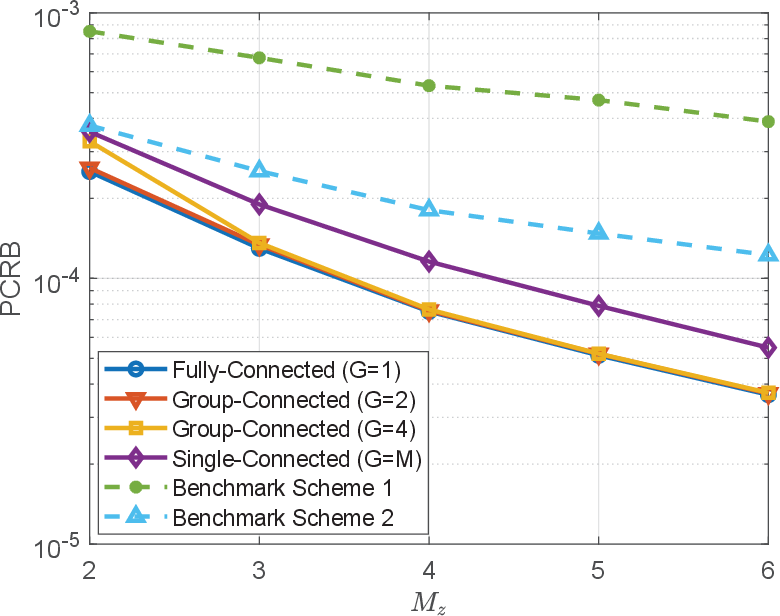}	\label{sensing_Mz}}
	\subfigure[PCRB versus transmit SNR.]{\includegraphics[width=.315\linewidth]{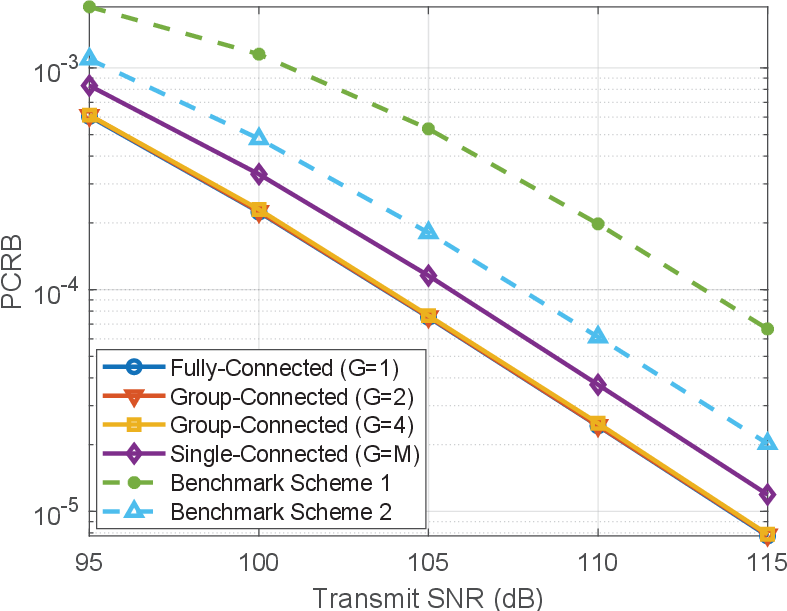} \label{PCRB_vsNoise}}
%		\vspace{-2mm}
	\caption{PCRB versus $M_x$/$M_z$/transmit SNR for BD-IRS aided sensing.}
	\vspace{-.5cm}
\end{figure*}

\begin{figure}[t]
	\centering
	\addtolength{\subfigcapskip}{-1mm}
	\setcounter{figure}{3}
	\subfigure[Convergence behavior for (P5).]{	\includegraphics[width=.473\linewidth]{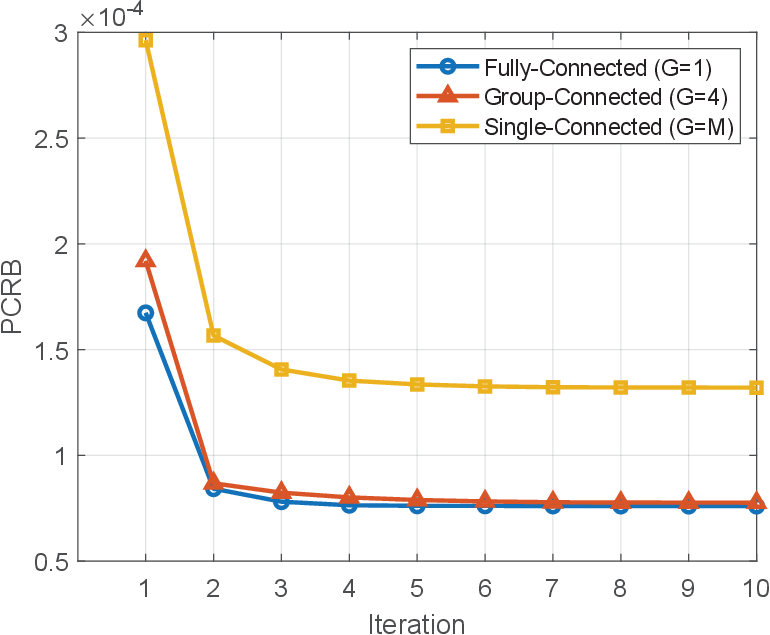}	\label{sensing_convergence}}
	\subfigure[Convergence behavior for (P1).]	{\includegraphics[width=.473\linewidth]{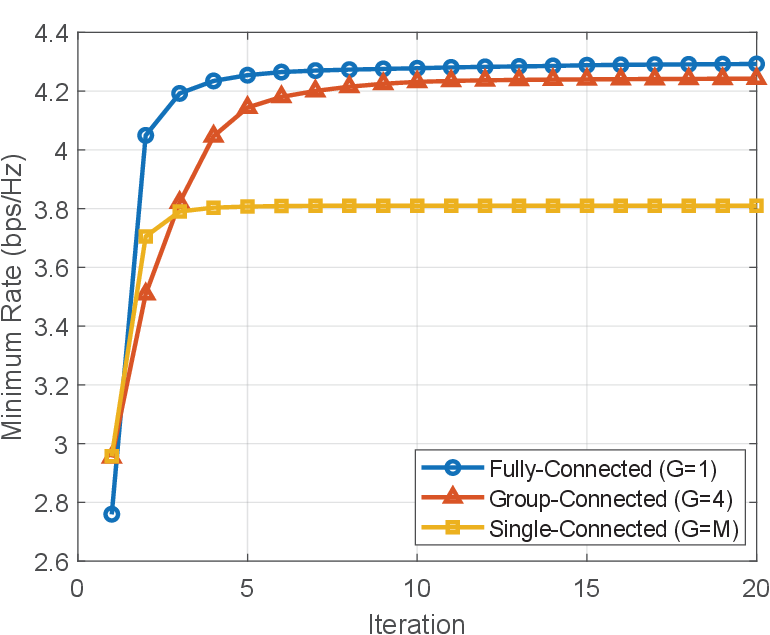} \label{ISAC_convergence}}
%			\vspace{-2mm}
	\caption{Convergence behavior of the proposed PDD-based algorithms for sensing-only and ISAC systems.}
	\vspace{-.5cm}
\end{figure}

 First, Fig. \ref{sensing_convergence} and Fig. \ref{ISAC_convergence} show the convergence behavior of the proposed PDD-based algorithms for the PCRB minimization problem in (P5) and the minimum rate maximization problem in (P1), where monotonic and quick convergence (within 10 and 20 iterations) is observed for both problems. This validates the effectiveness of the PDD-based algorithms.

\vspace{-.3cm}
\subsection{Evaluation of Sensing Performance Aided by BD-IRS}

 In this subsection, we evaluate the performance of sensing in terms of PCRB with the aid of BD-IRS when no communication user is present. Fig. \ref{sensing_Mx} and Fig. \ref{sensing_Mz} show the PCRB versus the number of reflecting elements in the $x$-axis and $z$-axis (i.e., $M_x$ and $M_z$) with the other one fixed as $4$, respectively. Fig. \ref{PCRB_vsNoise} shows the PCRB versus transmit SNR $\frac{P_0}{\sigma^2}=\frac{P_1}{\sigma^2}=\frac{P_2}{\sigma^2}$. It is observed from that the PCRB decreases as $M_x$, $M_z$, or transmit SNR increases; moreover, the fully-connected and group-connected architectures outperform the single-connected architecture (i.e., conventional diagonal IRS) as well as the two benchmark schemes, where the performance gain of BD-IRS over diagonal IRS enlarges as $M_x$ or $M_z$ increases. Additionally, it is observed that increasing $M_x$ is more effective in lowering the PCRB compared with increasing $M_z$, since the increment in $M_x$ can provide additional sensing information on $\theta$, which can be drawn from $\dot{\bm{g}}(\theta)$ in the PCRB expression, while increasing $M_z$ can mainly enhance the overall target-BS effective channel gain (or aperture gain). 
%\begin{table*}[t]
%	\centering
%	\caption{Performance and complexity comparison of fully-, group-, and single-connected BD-IRSs ($M_x=M_z=4$). }
%	%		\vspace{-.3cm}
%	\label{comparison}
%	\begin{tabular}{|c|c|c|c|c|c|c|c|}
%		\hline
%		Architecture & \makecell{Group\\Number} & \makecell{Circuit\\Topology\\Complexity} & \makecell{Optimization\\Computational\\Complexity} & \makecell{PCRB\\$(\times 10^{-5})$}  & \makecell{Performance\\Gain (PCRB)}& \makecell{Minimum\\Rate}&\makecell{Performance\\Gain \\(Minimum Rate)}\\
%		\hline
%		Fully-connected & $1$ & $\frac{M(M+1)}{2}$ & $\mathcal{O}(K^{1.5}M^6)$ & $7.506$ & $35.142\%$&$4.078$&$11.852\%$\\
%		\hline
%		Group-connected & $2$ & $\frac{M(M/2+1)}{2}$ & $\mathcal{O}(\frac{K^{1.5}M^6}{8})$ & $7.560$ & $34.680\%$&$4.044$&$10.912\%$\\
%		\hline
%		Group-connected & $4$ & $\frac{M(M/4+1)}{2}$ & $\mathcal{O}(\frac{K^{1.5}M^6}{64})$ & $7.642$ &$33.968\%$&$3.963$&$8.687\%$\\
%		\hline
%		Single-connected & $M$ & $M$ & $\mathcal{O}(K^{1.5}M^3)$ & $11.573$ &\diagbox{\hspace{.62cm}}{}&$3.646$&\diagbox{\hspace{.82cm}}{}\\
%		\hline
%	\end{tabular}
%	%	\vspace{-.2cm}
%\end{table*}

\begin{table}[t]
	\centering
	\caption{Performance and complexity comparison of fully-, group-, and single-connected BD-IRSs ($M_x=M_z=4$). }
			\vspace{-.2cm}
	\label{comparison}
	\begin{tabular}{|c|c|c|c|c|}
		\hline
		Architecture&Fully-&\multicolumn{2}{c|}{Group-} &Single-\\
		\hline
		\makecell{Group\\Number} & $1$ & $2$ & $4$ & $M$\\
		\hline
		\makecell{Circuit\\Topology\\Complexity}& $\!\frac{M(M+1)}{2}\!$ & $\!\frac{M(M/2+1)}{2}\!$ & $\!\frac{M(M/4+1)}{2}\!$ & $\!M\!$\\
		\hline
		\makecell{Optimization\\\!Computational\!\\Complexity} & $\!\!\suo\mathcal{O}(\!K^{1.5}\!M^6\!)\!\!\suo$ & $\!\!\suo\mathcal{O}(\!\frac{K^{1.5}\!M^6}{8}\!)\!\!\suo$ & $\!\!\suo\mathcal{O}(\!\frac{K^{1.5}\!M^6}{64}\!)\!\!\suo$ & $\!\!\suo\mathcal{O}(\!K^{1.5}\!M^3\!)\!\!\suo$\\
		\hline
		\makecell{PCRB\\$(\times 10^{-5})$} & $7.506$ & $7.560$ & $7.642$ & $11.573$\\
		\hline
		\makecell{Performance\\Gain (PCRB)} & $35.142\%$ & $34.680\%$ & $33.968\%$ &\diagbox{\hspace{.365cm}}{\vspace{.02cm}}\\
		\hline
		\makecell{Minimum\\Rate} & $4.078$ &$4.044$ &$3.963$ & $3.646$\\
		\hline
		\makecell{Performance\\Gain \\ \!\!(Minimum Rate)\!\!} &$11.852\%$ &$10.912\%$ &$8.687\%$ & \diagbox{\hspace{.365cm}}{\vspace{.2cm}}\\
		\hline
	\end{tabular}
		\vspace{-.5cm}
\end{table}

\vspace{-.4cm}
\subsection{Evaluation of ISAC Performance Aided by BD-IRS}
%\vspace{-.2cm}

\begin{figure*}[tbp]
	\centering
	\addtolength{\subfigcapskip}{-1.5mm}
	\setcounter{figure}{5}
	\subfigure[Minimum rate versus $\Gamma_{\mathrm{PCRB}}$.]{\includegraphics[width=.305\linewidth]{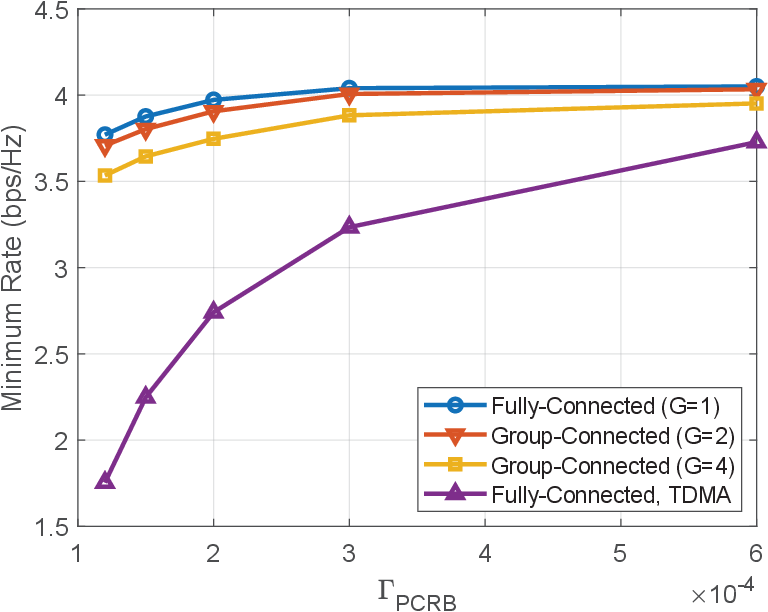} \label{ISAC_vsPCRB}}		
	\subfigure[Minimum rate versus $M_z$.]{\includegraphics[width=.305\linewidth]{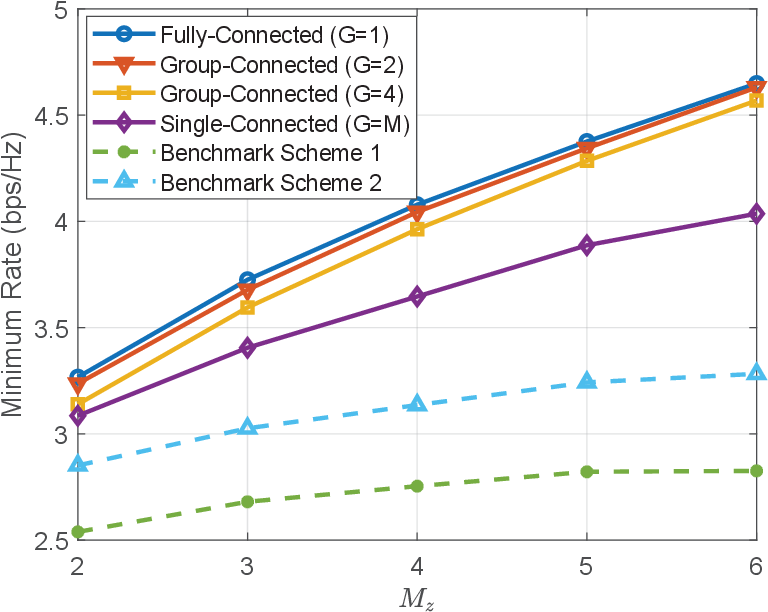} \label{ISACvsMz}}
	\subfigure[Minimum rate versus transmit SNR.]{\includegraphics[width=.305\linewidth]{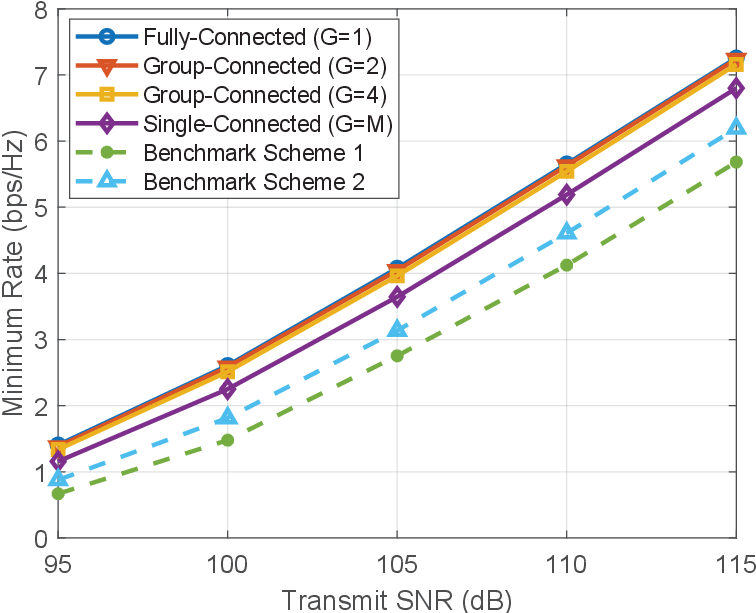} \label{ISACvsNoise}}	
	\vspace{-2mm}
	\caption{Minimum rate versus $\Gamma_{\mathrm{PCRB}}$/$M_z$/transmit SNR in BD-IRS aided uplink ISAC.}
	\vspace{-.5cm}
\end{figure*}

\begin{figure}[t]
	\centering
	
	\includegraphics[width=.76\linewidth]{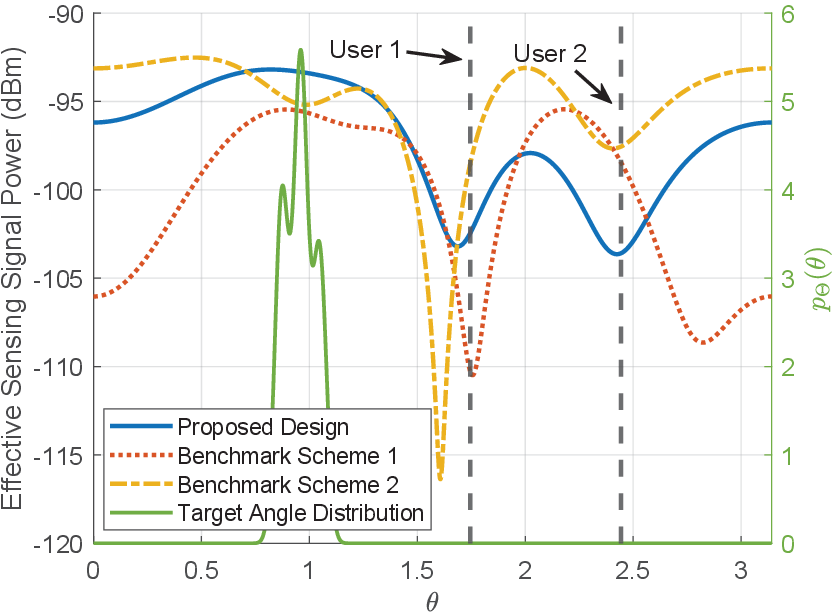}
		\vspace{-3.5mm}
	\caption{Effective sensing signal power received at the BS versus $\theta$.}
	\label{effective_sensing_power}
	%	\vspace{-5mm}
		\vspace{-.46cm}
\end{figure}

Next, we evaluate the performance of ISAC aided by BD-IRS. In Fig. \ref{ISAC_vsPCRB}, we show the minimum rate versus the PCRB threshold $\Gamma_{\mathrm{PCRB}}$. Benchmark schemes and the single-connected IRS are not shown for comparison as they may not always yield a feasible solution to Problem (P1). The results indicate that the minimum rate achieved by the proposed design increases as $\Gamma_{\mathrm{PCRB}}$ increases, which validates the trade-off between sensing and communication. Moreover, the TDMA scheme is generally outperformed by the SDMA scheme, due to the excessive time waste resulting from orthogonal signal transmission. Then, we set $\Gamma_{\mathrm{PCRB}}=1.5\times 10^{-3}$ and show the minimum rate versus $M_z$ in Fig. \ref{ISACvsMz}. It can be observed that the minimum rate increases as $M_z$ increases, and the proposed design outperforms both benchmark schemes by a large margin. Furthermore, it is observed that as the number of groups decreases, the performance improves at a cost of a more complicated architecture, which is summarized in Table \ref{comparison}. In addition, it is also noted that BD-IRS can use less elements to achieve the same performance as conventional IRS, which is helpful to reduce the hardware cost and physical size. For instance, in Fig. \ref{ISACvsMz}, BD-IRS with $2$ groups and $M_z=4$ can achieve comparable performance as conventional IRS with $M_z=6$.
In Fig. \ref{ISACvsNoise}, we show the minimum rate versus transmit SNR with $\Gamma_{\mathrm{PCRB}}=2.2\times 10^{-3}$. It is observed that the minimum rate increases as the transmit SNR increases. On the other hand, we show the effective sensing signal power in Fig. \ref{effective_sensing_power},  which is defined as the effective sensing signal power received at the BS from a location at distance $10$ m and angle $\theta$ with respect to the BD-IRS  via reflection of the BD-IRS, i.e., $P_0\vert \bm{w}_0^H \bm{R}\bm{\Phi}\bm{g}(\theta)\vert^2$, where $\bm{w}_0\in\mathbb{C}^{N\times 1}$ denotes an auxiliary receive beamforming vector at the BS for the sensing signal.\footnote{Note that the PCRB derived in this paper can bound the MSE with any receive signal processing and estimation method, while we consider a specific receive beamforming design here only for the purpose of illustrating the effect of BD-IRS reflection.}
Motivated by the relationship between the sensing SINR and the PCRB, $\bm{w}_0$ is designed to maximize the expected sensing SINR given by 
	\begin{align}
		\bar{\gamma}_\mathrm{s}(\bm{w}_0,\bm{\Phi})&\triangleq\mathbb{E}_\theta [\gamma_\mathrm{s}(\bm{w}_0,\bm{\Phi},\theta)]\nonumber\\
		&= \frac{P_{0}  \bm{w}_0^H \mathbb{E}_\theta[\bm{R}\bm{\Phi}\bm{g}(\theta)\bm{g}^H(\theta)\bm{\Phi}^H\bm{R}^H]\bm{w}_0}{\bm{w}_0^H\bm{\Sigma}_0\bm{w}_0},
	\end{align}
	which is the eigenvector corresponding to the largest eigenvalue of $\bm{\Sigma}_0^{-1} \mathbb{E}_\theta [(\bm{R}\bm{\Phi}\bm{g}(\theta))(\bm{R}\bm{\Phi}\bm{g}(\theta))^H]$.
It can be observed that our proposed design concentrates power towards the angles with high probabilities while suppressing power in the directions of communication users to reduce interference.
 This validates the effectiveness of our proposed design as well as its gain over the benchmark schemes which cannot achieve the above location-dependent interference management.

\vspace{-.3cm}
\subsection{Comparison of TDMA Versus SDMA}

\begin{figure}[t]
	\centering
	\includegraphics[width=.7\linewidth]{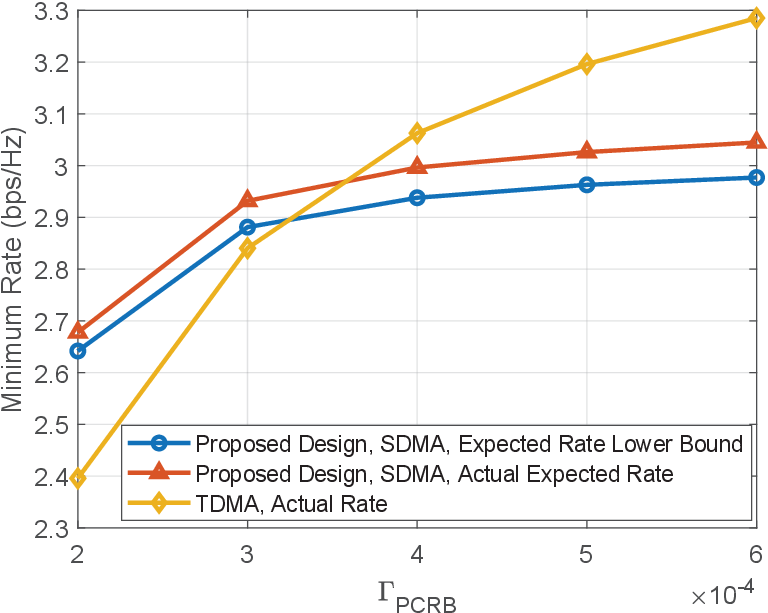}
	\vspace{-3.5mm}
	\caption{Comparison between TDMA and SDMA under severe interference.}
	\label{ISAC_TDMA_good}
	\vspace{-.45cm}
\end{figure}

Finally, we consider a scenario with severe interference between sensing and communication, to reveal the potential gain of TDMA over SDMA. Specifically, we set $\theta^{\mathrm{U},1}=\frac{5}{12}\pi$, where the direct user-BS channel is blocked. Note that in this case, user 1 is very close to the highly-probable locations of the target, thereby leading to significant interference. Fig. \ref{ISAC_TDMA_good} compares the performance of SDMA via our proposed algorithm for Problem (P1) and TDMA, where both the actual expected rate of SDMA and its approximate lower bound are shown. It is observed that as the sensing requirement becomes less stringent, i.e., $\Gamma_{\mathrm{PCRB}}$ increases, TDMA outperforms SDMA. In this case, more time can be allocated to communication, and the pre-log factor in the TDMA communication rate can be compensated by the larger SINR achieved by removing the interference from sensing. This demonstrates the effectiveness of TDMA in heavy-interference scenarios. In addition, it is observed that the actual expected rate and its lower bound for SDMA are very close, which validates our adoption of the lower bound as the communication performance metric.

\vspace{-.1cm}
\section{Conclusions}\label{sec7}\vspace{-.1cm}
This paper studied the optimization of BD-IRS reflection in a BD-IRS aided multi-user uplink ISAC system, where the BS serves multiple uplink users and senses the unknown and random location information of an active target via its sent uplink signals and the location PDF information. The PCRB was firstly derived to characterize the sensing performance and then transformed to a tractable form. Following this, a BD-IRS reflection matrix optimization problem was formulated to maximize the minimum expected rate among multiple users while satisfying a constraint on the PCRB as well as the structural constraints on the reflection matrix. The proposed problem was challenging to solve since it included various types of non-convex components. To address this challenging problem, we proposed a PDD-based algorithm which yields a high-quality suboptimal solution with polynomial-time complexity. We also proposed an alternative TDMA scheme with optimized time allocation. Numerical results validated the effectiveness of the proposed designs and provided useful design insights such as the optimal choice of multiple access scheme.
In the future, it is worthwhile extending this work to other emerging new IRS architectures such as active IRS and stacked IRS \cite{Rtwo}, as well as to cases with joint angle and range estimation and/or multiple sensing targets.

\begin{appendix}
	\subsection{Proof of Proposition 1}\label{app1}
	Define 
	\vspace{-.1cm}\begin{align}
	\bar{\gamma}_k\suo(\suo\bm{w}_k,\!\bm{\Phi}\suo)\!\triangleq\!\frac{P_{k} \vert \bm{w}_k^H \bm{h}_k(\bm{\Phi})\vert^2}{\suo\sum\limits_{\substack{k'=1\\k' \neq k}}^K \! P_{k'} \!\vert \suo\bm{w}_k^H \suo\bm{h}_{k'}\suo(\suo\bm{\Phi}\suo) \vert^2 \suo\suo+\suo\suo  P_0\suo\Vert   \suo\bm{w}_k^H \suo\bm{R}\bm{\Phi}\bm{G}^{\frac{1}{2}} \suo\Vert^2 \suo\suo+\suo\suo \Vert \suo\bm{w}_k\suo\Vert^2\sigma^2}\nonumber\\[-12pt]
	\end{align}
	 and consequently $\bar{R}_k(\bm{w}_k,\bm{\Phi})=\log_2(1+\bar{\gamma}_k(\bm{w}_k,\bm{\Phi}))$ holds.
	Since $\bar{R}_k(\bm{w}_k,\bm{\Phi})$ is a monotonically increasing function of $\bar{\gamma}_k(\bm{w}_k,\bm{\Phi})$, rate maximization is equivalent to SINR maximization. Since $\bar{\gamma}_k(\bm{w}_k,\bm{\Phi})$ is independent on $\{\bm{w}_{k'},k'\neq k\}_{k'=1}^K$, 
	 the optimization problem of $\bm{w}_k$ is given by
	\begin{align}
		\mbox{(P-$\bm{w}_k$)}
		\quad\max_{\bm{w}_k}&\quad \frac{P_{k} \vert \bm{w}_k^H \bm{h}_k(\bm{\Phi})\vert^2}{\bm{w}_k^H \bm{\Sigma}_k(\bm{\Phi}) \bm{w}_k},
	\end{align}
	where $\bm{\Sigma}_k(\bm{\Phi}) = \sum_{k'=1,k' \neq k}^K P_{k'} \bm{h}_{k'}(\bm{\Phi})\bm{h}_{k'}^H(\bm{\Phi})
	+P_0  \bm{R}\bm{\Phi}\bm{G} \bm{\Phi}^H\bm{R}^H
	+\sigma^2\bm{I}_N$.
	Denote $\bm{z}_k=\bm{\Sigma}_k^{\frac{1}{2}}(\bm{\Phi})\bm{w}_k$. Thus, we can express objective of Problem (P-$\bm{w}_k$) as 
	\begin{align}
		\max_{\bm{z}_k}\quad	\frac{P_{k} \bm{z}_k^H\bm{\Sigma}_k^{-\frac{1}{2}}(\bm{\Phi})\bm{h}_k(\bm{\Phi}) \bm{h}_k^H(\bm{\Phi})\bm{\Sigma}_k^{-\frac{1}{2}}(\bm{\Phi})\bm{z}_k}{\bm{z}_k^H \bm{z}_k}.
	\end{align}
	This problem is a Rayleigh quotient maximization problem and the optimal $\bm{z}_k^\star$ should align with the eigenvector corresponding to the largest eigenvalue of  $\bm{\Sigma}_k^{-\frac{1}{2}}(\bm{\Phi})\bm{h}_k(\bm{\Phi})\bm{h}_k^H(\bm{\Phi})\bm{\Sigma}_k^{-\frac{1}{2}}$ \cite{horn2012matrix}. Therefore, the optimal solution is given by $\bm{z}_k^\star = \bm{\Sigma}_k^{-\frac{1}{2}}(\bm{\Phi})\bm{h}_k(\bm{\Phi})$. Then, we have $\bm{w}_k=\bm{\Sigma}_k^{-\frac{1}{2}}(\bm{\Phi})\bm{z}_k^\star=\bm{\Sigma}_k^{-1}(\bm{\Phi})\bm{h}_k(\bm{\Phi})$. Normalizing $\bm{w}_k$, we obtain the optimal $\bm{w}_k^\star$ in \eqref{wopt}.
	\subsection{Proof of Proposition \ref{pro2}}\label{app2}
	$f_{\zeta}(\bm{\Phi},\bm{\nu}_{\zeta})$ can be rewritten as
	\begin{align}
		&f_{\zeta}(\bm{\Phi},\bm{\nu}_{\zeta})=\bm{\nu}_{\zeta}^H\bm{\Sigma}_0(\bm{\Phi})\bm{\nu}_{\zeta}-2\mathfrak{Re}\{\bm{\nu}_{\zeta}^H\bm{R\Phi u}_{\zeta}\}\\
		&\hspace{.25cm}=\big(\bm{\nu}_{\zeta}-\bm{\Sigma}_0^{-1}(\bm{\Phi})\bm{R\Phi u}_{\zeta}\big)^H  \bm{\Sigma}_0(\bm{\Phi}) \big(\bm{\nu}_{\zeta}-\bm{\Sigma}_0^{-1}(\bm{\Phi})\bm{R\Phi u}_{\zeta}\big)\nonumber\\
		&\hspace{.85cm}-(\bm{R\Phi u}_{\zeta})^H \bm{\Sigma}_0^{-1}(\bm{\Phi}) (\bm{R\Phi u}_{\zeta})\label{proof1}.
	\end{align}
	The optimal $\bm{\nu}_{\zeta}^\star$ that minimizes $f_{\zeta}(\bm{\Phi},\bm{\nu}_{\zeta})$ is given by
	\begin{gather}
		\bm{\nu}_{\zeta}^\star=\bm{\Sigma}_0^{-1}(\bm{\Phi})\bm{R\Phi u}_{\zeta}.\label{proof2}
	\end{gather}
	Then, we have
	\begin{align}
		\!\!\suo f_{\zeta}(\bm{\Phi},\bm{\nu}_{\zeta}^\star)
		&	\!=\!-(\bm{R\Phi u}_{\zeta})^H\bm{\Sigma}_0^{-1}(\bm{\Phi})(\bm{R\Phi u}_{\zeta}),\\
		\!\!\suo\sum\limits_{\zeta=1}^{R} \kappa_{\zeta} f_{\zeta}(\bm{\Phi},\bm{\nu}_{\zeta}^\star)&\!=\!	-\sum\limits_{\zeta=1}^{R} \suo\kappa_{\zeta} (\bm{R\Phi u}_{\zeta})^H\bm{\Sigma}_0^{-1}\suo(\bm{\Phi})\suo(\bm{R\Phi u}_{\zeta}).\!\!\!\!\label{appd2}
	\end{align}
Similarly, it can be shown that the optimal $\bm{\nu}_k^\star$ to minimize each $f_k(\bm{\Phi},\bm{\nu}_k)$ is given by $\bm{\nu}_k^\star=\bm{\Sigma}_k^{-1}(\bm{\Phi})\bm{h}_k(\bm{\Phi})$. $f_k(\bm{\Phi},\bm{\nu}_k^\star)$ can be further derived as $-\bm{h}^H_k(\bm{\Phi})\bm{\Sigma}_k^{-1}(\bm{\Phi})\bm{h}_k(\bm{\Phi})$. The equivalence between Problem (P4-I) and Problem (P4-I-eqv) thus follows by plugging in the values of $f_{\zeta}(\bm{\Phi},\bm{\nu}_{\zeta}^\star)$ and $f_k(\bm{\Phi},\bm{\nu}_k^\star)$ into Problem (P4-I-eqv).
\subsection{Construction of $\bm{Q}_g$}\label{Qg}
	 For a given ${\Phi_g}_{m,n}$, the index of ${\Phi_g}_{m,n}$ should appear at the  $i_2=(n-1)M_g+m$-th position of $\bm{\varphi}_g$. Moreover, we define $\widehat{M}_g\triangleq\sum_{g'=1}^{g-1} M_{g'}$ and $\widehat{M}_1\triangleq0$. Therefore, the index of ${\Phi_g}_{m,n}$ in $\mathrm{vec}(\bm{\Phi})$ is given by $i_1=(\widehat{M}_g+n-1)M+\widehat{M}_g+m$. Thus, the $((\widehat{M}_g+n-1)M+\widehat{M}_g+m,(n-1)M_g+m)$-th entry of $\bm{Q}_g$ should be $1$. For all $(m,n)$ pairs, we can calculate the corresponding $(i_1,i_2)$. By denoting the collection of all $(i_1,i_2)$ pairs as $\mathcal{I}$, the expression of $\bm{Q}_g$ is given by ${Q_{g}}_{i_1,i_2}=1$ if $(i_1,i_2) \in \mathcal{I}$ and ${Q_{g}}_{i_1,i_2}=0$ otherwise.
%	\begin{align}
%		{Q_{g}}_{i_1,i_2}
%		\begin{cases}
%			1, &\text{if } (i_1,i_2) \in \mathcal{I}\\
%			0, &\text{otherwise}.
%		\end{cases}
%	\end{align}
\subsection{Construction of $\bm{D}_g$}\label{Dg}
	The explicit form of $\bm{D}_g$ can be expressed as 
	$\bm{D}_g=\big(\sum_{m\geq n}\bm{e}_{mn}(\mathrm{vec}(\bm{T}_{mn}))^T\big)^T$ \cite{duplication}, where $\bm{e}_{mn}\in\mathbb{R}^{\frac{M_g(M_g+1)}{2}\times 1},1\leq n\leq m\leq M_g$ is a unit vector with element $1$ in the $[(n-1)M_g+m-\frac{1}{2}n(n-1)]$-th position and zeros elsewhere, and $\bm{T}_{mn}\in\mathbb{R}^{M_g \times M_g}$ is a matrix with $1$ in the $(m,n)$-th and $(n,m)$-th position and zeros elsewhere, respectively. 
\end{appendix}

\bibliographystyle{IEEEtran}
\bibliography{IEEEabrv,reference}

\end{document}